\documentclass[prl,aps,twocolumn,amsfonts,showpacs,superscriptaddress,preprintnumbers,nofootinbib]{revtex4-1}
 
\usepackage{graphicx}
\usepackage{xcolor}
\usepackage{rotating}
\usepackage{bm,amsmath,amssymb}
\usepackage[utf8]{inputenc}
\usepackage{footmisc}

\usepackage[colorlinks=true, urlcolor=blue, linkcolor=blue, citecolor=blue, hyperindex=true, linktocpage=true]{hyperref}

\newcommand{\nn}{\nonumber\\}

\def\CC{\mathcal{C}}

\def\CN{\mathcal{N}}

\def\re{\mbox{Re}}
\def\im{\mbox{Im}}

\def\q{{\bf q}}

\def\be{\begin{equation}}
\def\ee{\end{equation}}

\def\bea{\begin{eqnarray}}
\def\eea{\end{eqnarray}}

\def\im{\text{Im}}
\def\re{\text{Re}}

\begin{document}
\preprint{MIT-CTP/5226}

\title{Bounds on transport from univalence and pole-skipping}

\author{Sa\v{s}o Grozdanov}
\affiliation{Center for Theoretical Physics, MIT, Cambridge, MA 02139, USA}
\affiliation{Faculty of Mathematics and Physics, University of Ljubljana, Jadranska ulica 19, SI-1000 Ljubljana, Slovenia}

\begin{abstract}
Bounds on transport represent a way of understanding allowable regimes of quantum and classical dynamics. Numerous such bounds have been proposed, either for classes of theories or (by using general arguments) universally for {\em all} theories. Few are exact and inviolable. I present a new set of methods and sufficient conditions for deriving exact, rigorous, and sharp bounds on all coefficients of hydrodynamic dispersion relations, including diffusivity and the speed of sound. These general techniques combine analytic properties of hydrodynamics and the theory of univalent (complex holomorphic and injective) functions. Particular attention is devoted to bounds relating transport to quantum chaos, which can be established through pole-skipping in theories with holographic duals. Examples of such bounds are shown along with holographic theories that can demonstrate the validity of the conditions involved. I also discuss potential applications of univalence methods to bounds without relation to chaos, such as for example the conformal bound on the speed of sound.
\newline
\end{abstract}
\maketitle

{\bf Introduction.---}The existence of bounds on properties of transport, such as diffusion, has persistently enthralled physicists concerned with time-dependent collective dynamics. Numerous bounds that improved our understanding of quantum and classical dynamics have been proposed. Among them is Sachdev's relaxation time bound \cite{sachdev_2011}, the Mott-Ioffe-Regel limit of metallic conductivity \cite{ioffe1960non,mott1972conduction}, lower bounds on diffusion and viscosity \cite{Kovtun:2004de,Hartnoll:2014lpa,Blake:2016wvh,Zaanen:2018edk,trachenko2020minimal,Baggioli:2020ljz,Kovtun:2011np,Chafin:2012eq,Kovtun:2014nsa,Martinez:2017jjf}, upper bounds on diffusion \cite{Lucas:2016yfl,Hartman:2017hhp,Baggioli:2020ljz} and a bound on the speed of sound \cite{Cherman:2009tw,Hohler:2009tv}. These bounds are usually heuristic and rely on basic physical principles such as the uncertainty principle and causality. Exact inequalities, even for restricted classes of theories are rare. An example is Prosen's bound on diffusion \cite{Prosen_2014}. Holographic methods to bound conductivities in disordered theories were developed in \cite{Grozdanov:2015qia,Grozdanov:2015djs}. Holographic advances in quantum chaos then led to the exact Maldacena-Shenker-Stanford bound on quantum Lyapunov exponents that follows from arguments of analyticity and complex analysis \cite{Maldacena:2015waa}. Another bound on the growth of weak (polynomial) quantum chaos was derived in \cite{Kukuljan:2017xag}.

Microscopic bounds, such as bounds on quantum chaos, should imply sharp bounds on collective transport. The purpose of this work is to introduce a new set of mathematical techniques from a well-developed theory of univalent functions, which allows for a rigorous derivation of exact inequalities of that type on diffusivity, the speed of sound and all higher-order coefficients of hydrodynamic dispersion relations. The methods establish sufficient analyticity and microscopic conditions that lead to several long-discussed types of bounds. Due to their generality, univalence methods can also be applied to derive bounds without any reference to chaos.
 
{\bf Univalent functions.---}A univalent (or {\em schlicht}) function $f(z)$ is a complex holomorphic injective function. The condition of injectivity demands that $f(z_1) \neq f(z_2)$ for all $z_1 \neq z_2$. Henceforth, all considered $f(z)$ will be univalent in some simply connected region $U\subset \mathbb{C}$. By the {\em Riemann mapping theorem}, it is then possible to map $U$ to an open unit disk $\mathbb{D} = \{\zeta\, | \, |\zeta| < 1 \}$ in the complex $\zeta$-plane by a holomorphic invertible conformal map $\varphi$: $\zeta = \varphi(z)$ and $z = \varphi^{-1}(\zeta)$. As is conventional, we will use the normalisation $f(\zeta = 0) = 0$, and $f'(\zeta = 0) = 1$ for functions in the $\zeta$-plane. All such functions admit a power series representation of the following form:
\begin{equation}\label{UnivFunc}
f(\zeta) = \zeta + \sum_{n=2}^\infty b_n \zeta^n.
\end{equation}
The series is guaranteed to converge for all $|\zeta| < 1$.

Locally, $f(z)$ is univalent if $f'(z) \neq 0$. However, proving local univalence at every $z \in U$ does not guarantee global univalence. Instead, one of numerous sufficient conditions for univalence must be employed \cite{duren2010univalent,lehto2011univalent}. Once univalence is established and we have mapped $U \to \mathbb{D}$, then we can resort to theorems bounding univalent functions on $\zeta\in\mathbb{D}$, such as the {\em growth theorem}: 
\begin{equation}\label{GrowthTheorem}
\frac{|\zeta|}{\left( 1 + |\zeta|\right)^2} \leq \left|f(\zeta)\right| \leq \frac{|\zeta|}{\left( 1 - |\zeta|\right)^2} ,
\end{equation}
and the celebrated {\em de Branges theorem}  (originally called the {\em Bieberbach conjecture}) \cite{branges1985} constraining each coefficient of the power series \eqref{UnivFunc}:
\begin{equation}\label{deBranges}
| b_n | \leq n,\quad\text{for all}~n \geq 2. 
\end{equation}
The inequalities in Eq.~\eqref{deBranges} and the growth theorem \eqref{GrowthTheorem} are saturated by the Koebe function (and its rotations in $\zeta$),
\begin{equation}\label{Koebe}
f_K(\zeta) = \frac{\zeta}{(1-\zeta)^2} = \sum_{n=1} n \zeta^n,
\end{equation}
which conformally maps $\mathbb{D} \to \mathbb{C} \setminus (-\infty, -1/4]$.  

We will use the condition whereby if $\re \,f'(z) > 0 $ in any convex $U\subset\mathbb{C}$, then $f(z)$ is univalent in $U$ \cite{noshiro1934theory,warschawski1935higher}. If, moreover, after $\varphi:U\to\mathbb{D}$, $\re \, f'(\zeta) > 0$, then $f(\zeta)$ satisfies stronger versions of the theorems in Eqs.~\eqref{GrowthTheorem} and \eqref{deBranges} \cite{macgregor1962functions}:
\begin{align}
-|\zeta| + 2 \ln\left(1 + |\zeta|\right) &\leq |f(\zeta)| \leq -|\zeta| - 2\ln \left(1- |\zeta|\right) , \label{MacGregor1} \\
| b_n | &\leq \frac{2 }{ n},\quad\text{for all}~n \geq 2. \label{MacGregor2}
\end{align}

{\bf Hydrodynamics.---}Hydrodynamics is an effective theory of collective late-time and long-range excitations in fluids governed by conserved quantities such as energy, momentum, and charges \cite{landau,Kovtun:2012rj,Dubovsky:2011sj,Grozdanov:2013dba,Crossley:2015evo,Glorioso:2017fpd,Haehl:2015foa,Haehl:2015uoc,Jensen:2017kzi,Glorioso:2018wxw,Grozdanov:2016tdf,Chen-Lin:2018kfl}. Linearised hydrodynamics predicts the structure of dispersion relations $\omega(\q^2)$, where $\omega$ is the frequency and $\q^2$ the momentum (squared) of a collective mode: diffusion or sound. In theories preserving spatial rotations, classical\footnote{Classical hydrodynamics is a theory without stochastic noise or loop corrections leading to the breakdown of analyticity \cite{Kovtun:2003vj,Chen-Lin:2018kfl,Delacretaz:2020nit}.} $\omega(\q^2)$ are infinite Puiseux series in $\q^2$ \cite{Grozdanov:2019kge,Grozdanov:2019uhi}: 
\begin{align}
\omega_{\rm diff} (z \equiv {\bf q}^2) &= - i \sum_{n=1}^\infty c_n z^n  , \label{WDiff}\\
\omega^\pm_{\rm sound}  (z \equiv \sqrt{\q^2}) &= - i \sum_{n=1}^\infty a_n e^{\pm \frac{i\pi n}{2}} z^n ,  \label{WSound}
\end{align}
where all $a_n,c_n \in \mathbb{R}$. We treat the argument $z$ as complex ($z\in\mathbb{C}$) in both Eqs.~\eqref{WDiff} and \eqref{WSound}. We have $c_1 = D$ (diffusivity) and $a_1 = v_s$ (the speed of sound). Each series converges for $|z| < R \equiv |z_*|$ with $z = z_*$ being the first critical point of the associated complex curve \cite{Grozdanov:2019kge,Grozdanov:2019uhi}.\footnote{See also Refs. \cite{Withers:2018srf,Heller:2020uuy} and \cite{Abbasi:2020ykq,1809177}.} Each fully analytically continued function $\omega(z)$ is holomorphic in the region $z\in H \subset \mathbb{C}$, where $H$ contains $|z| < R$.

Different concepts of wave propagation speeds beyond $v_s$ exist, such as the phase velocity $v_{ph}(q) \equiv \omega / q$, the front velocity, and the group velocity $v_g(q) \equiv \partial \omega / \partial q$, where $q \equiv \sqrt{\q^2}$. Causality, for example, imposes certain conditions on these speeds (see Ref. \cite{krotscheck1978causality}). In an analogous spirit, we will sometimes use properties of $v_g$ to define the univalence region of hydrodynamics $U$.

{\bf General bounds.---}A hydrodynamic dispersion relation $\omega(z)$ is by Puiseux's theorem invertible at $z=0$ and thus locally univalent at $z=0$ \cite{Grozdanov:2019kge,Grozdanov:2019uhi}. Beyond including $z=0$ in all univalent regions $U \subseteq H$, we assume that $U$ also contains a point $z = z_0$ where $\omega_0 \equiv \omega(z_0)$ is known.  $U$ need not be maximal. A convenient way to choose $U$ is through the sufficient condition $\re\, f'(z) > 0$, where $f_{\rm diff}(z) = i \omega_{\rm diff}(z)$ and $f_{\rm sound}(z) = \omega_{\rm sound}(z)$. This implies univalence for $U = \{z  \, | \, |z| < \min[|z_g| , R]\}$, where 
\begin{align}
\!\!\! {\rm diffusion}:&~ z_g = q_g^2 \equiv \min q^2 \, | \, \re\,v_g \,\im\,q = \im\,v_g \,\re\,q, \label{GroupVelocityConditionShear} \\
{\rm sound}:&~ z_g = q_g \equiv \min q \, | \,  \re\,v_g = 0, \label{GroupVelocityConditionSound}
\end{align}
expressed through the properties of the group velocity. If $v_g$ vanishes at $|z_g|$ smaller than those in \eqref{GroupVelocityConditionShear} and \eqref{GroupVelocityConditionSound}, then univalence is lost locally due to $f'(z_g) = 0$. We have
\begin{equation}
q_g\equiv \min q \, | \, v_g = 0 .  \label{GroupVelocityCondition}
\end{equation}

Using a conformal map $\varphi:U \to \mathbb{D}$ with $\varphi(z) = \zeta$ that preserves the origin (i.e., $\varphi(0) = 0$), we then define
\begin{align}
f_{\rm diff}(\zeta) &\equiv \frac{i \omega_{\rm diff}(\varphi^{-1}(\zeta) ) }{D \partial_\zeta \varphi^{-1}(0)} = \zeta + \sum_{n=2}^\infty b^{\rm diff}_n \zeta^n, \label{FDiff}\\ 
f_{\rm sound}(\zeta) &\equiv \frac{\omega^+_{\rm sound}(\varphi^{-1}(\zeta) ) }{v_s \partial_\zeta \varphi^{-1}(0)} = \zeta + \sum_{n=2}^\infty b^{\rm sound}_n \zeta^n. \label{FSound}
\end{align}
Both Eqs.~\eqref{FDiff} and \eqref{FSound} have the form of Eq.~\eqref{UnivFunc}. The growth theorem \eqref{GrowthTheorem} applied at $\zeta_0 \equiv \varphi(z_0)$ now yields lower and upper bounds on diffusivity and the speed of sound: 
\begin{equation}\label{DBoundGeneral}
 \frac{\left| \omega_0 \right| \left( 1 - |\zeta_0|\right)^2  }{|\zeta_0| \left| \partial_\zeta \varphi^{-1}(0) \right| } \leq ( D \lor v_s) \leq \frac{ \left| \omega_0 \right| \left( 1 + |\zeta_0|\right)^2  }{|\zeta_0| \left| \partial_\zeta \varphi^{-1}(0) \right|} ,
\end{equation}
where $( D \lor v_s)$ means either $D$ or $v_s$ depending on whether we used Eq.~\eqref{FDiff} or Eq.~\eqref{FSound}. If, in addition to univalence, $\re \, f'(\zeta) > 0$ for $|\zeta| < 1$, then Eq.~\eqref{MacGregor1} gives
\begin{align}\label{DBoundGeneralReal}
 & \frac{\left| \omega_0 \right| }{\left| \partial_\zeta \varphi^{-1}(0) \right|\ln  \left[ e^{-|\zeta_0|} / \left(1 - |\zeta_0|\right)^2 \right] }   \leq ( D \lor v_s) \nn
 &\leq  \frac{\left| \omega_0 \right|  }{\left| \partial_\zeta \varphi^{-1}(0) \right| \ln \left[ e^{-|\zeta_0|} \left(1 + |\zeta_0|\right)^2 \right] }  .
\end{align}
To bound higher-order coefficients, we use the de Branges theorem \eqref{deBranges} on each term of the series \eqref{FDiff} or \eqref{FSound}. This establishes a chain of inequalities on $c_n$ or $a_n$ in terms of all $c_m$ or $a_m$ with $m < n$. For a diffusive dispersion relation \eqref{WDiff}, we first use $|b_2| \leq 2$ to bound $c_2$:
\begin{equation}\label{C2Bound}
\left| c_2 + \frac{D}{2} \frac{ \partial^2_\zeta \varphi^{-1}(0) }{ \left[ \partial_\zeta \varphi^{-1}(0) \right]^2} \right| \leq \frac{2 D}{\left| \partial_\zeta \varphi^{-1}(0)  \right|},
\end{equation}
further eliminating $D$ through Eq.~\eqref{DBoundGeneral}. Next, $|b_3| \leq 3$ is used to bound $c_3$ and so on for all $c_{n\geq 4}$. If $\re\,f'(\zeta) > 0$, then the bound \eqref{C2Bound} has another factor of $1/2$ on the right-hand-side due to $|b_2| \leq 1$ in Eq.~\eqref{MacGregor2}. An analogous procedure can be used for bounding $a_n$ by $v_s$ and $\varphi$. All bounds are determined purely in terms of a single known $\omega_0(z_0)$ and the chosen original region of univalence $U$ through the conformal map $\varphi:U\to\mathbb{D}$. 

{\bf Quantum chaos and pole-skipping.---}Of particular interest are bounds that stem from the underlying microscopic quantum chaos. While the general relation between transport and chaos is unknown, precise connection has been established through the phenomenon of {\em pole-skipping} in quantum field theories with a large number of local degrees of freedom (large-$N$ theories) that possess a classical gravitational holographic dual \cite{Grozdanov:2017ajz,Blake:2017ris,Blake:2018leo,Grozdanov:2018kkt}. 

Pole-skipping is an indeterminacy of two-point functions associated with dispersion relations \eqref{WDiff}--\eqref{WSound}. In the longitudinal channel of energy-momentum fluctuations (e.g. sound or energy diffusion), pole-skipping implies 
\begin{equation}\label{PS-Long}
\omega_0(\q_0^2) = i \lambda_L, \quad  \q_0^2 = - \lambda_L^2 / v_B^2.
\end{equation}
Hence, for such modes, we have $q_0 = i \lambda_L / v_B$. Here, $\lambda_L$ is the maximal Lyapunov exponent $\lambda_L = 2\pi T$, $T$ is the temperature, and $v_B$ is the butterfly velocity characterising the exponential growth of the out-of-time-ordered correlator used to probe chaos $e^{\lambda_L (t - |{\bf x}|/v_B)}$ \cite{Shenker:2013pqa,Maldacena:2015waa}. In neutral theories, a related expression exists also for transverse fluctuations (e.g. momentum diffusion) \cite{Grozdanov:2019uhi,Blake:2019otz}:
\begin{equation}\label{PS-Trans}
\omega_0(\q_0^2) = - i \lambda_L, \quad  \q_0^2 =  \lambda_L^2 / v_B^2.
\end{equation}
In charged theories, pole-skipping in Eq.~\eqref{PS-Trans} at $\omega_0 = - i \lambda_L$ generically exhibits a more complicated $q_0$.\footnote{\label{Foot1}In general, the dispersion relations which give rise to Eqs.~\eqref{PS-Long} and \eqref{PS-Trans} also pass through an infinite sequence of pole-skipping points $\omega(\q_n^2) = - 2\pi T i n $ for all $n\in\mathbb{Z}_+ \cup \{0\}$ \cite{Grozdanov:2019uhi,Blake:2019otz}.} Since the pole-skipping points can be easily computed from dual gravity, and they relate chaos to transport, we will use them as $\omega_0(z_0)$ in most bounds below.

{\bf Diffusion I: Maximal univalence.---}In our first, simple, and very special example, assume that a diffusive dispersion relation $\omega(z)=\omega_{\rm diff}(z)$ (cf. Eq.~\eqref{WDiff}) is maximally univalent ($U = H$) and holomorphic on the entire $z\in\mathbb{C}$ except at a branch point $z_*$ and at $z = \infty$. We define $\omega_* \equiv \omega(z_*)$. Under $\im\,z \to - \im\,z$, $\re\,\omega$ is odd and $\im\,\omega$ is even. To have a single $z_*$, we need $\re\,\omega_* = 0$; hence $z_*\in \mathbb{R}$. For concreteness, we take $z_* > 0$ and choose the branch cut so that $U = \mathbb{C} \setminus [z_*,\infty)$. $R = z_*$ is the radius of convergence of the hydrodynamic series \eqref{WDiff}. We first use a rescaling M\"{o}bius transformation to map $z_* \to - 1/4$, keeping $z = \infty$ at $\infty$.  The branch cut is now chosen to lie along $(-\infty,-1/4]$. Next, we use an inverse of the Koebe function \eqref{Koebe} to map $\mathbb{C} \setminus (-\infty, -1/4] \to \mathbb{D}$. The full conformal map $\varphi:U\to\mathbb{D}$ is thus
\begin{align}
\zeta &= \varphi(z) = \frac{z - 2 z_* + 2 \sqrt{z_*^2- z z_* }}{z}, \\
z &= \varphi^{-1} (\zeta) = - 4 z_* f_K(\zeta) = - \frac{4 z_* \zeta}{(1- \zeta)^2 }, \label{GenDiffKoebe}
\end{align}
with $\partial^n_\zeta \varphi^{-1}(0) = - 4 n^2 (n-1)! R$. Using the pole-skipping relations in Eq.~\eqref{PS-Long} or Eq.~\eqref{PS-Trans}, the diffusivity bounds \eqref{DBoundGeneral} become
\begin{align}
z_0 = - \frac{\lambda_L^2}{v_B^2} < 0:&\quad \frac{v_B^2}{\lambda_L} \leq D \leq \frac{v_B^2}{\lambda_L} + \frac{\lambda_L}{R} , \label{SoundKoebe}\\
0 < z_0 = \frac{\lambda_L^2}{v_B^2} < R :&\quad  \frac{v_B^2}{\lambda_L} - \frac{\lambda_L}{R} \leq D \leq \frac{v_B^2}{\lambda_L} . \label{ShearKoebe}
\end{align}
Since $[z_*, \infty) \notin U$, we do not consider $z_0 \geq R$. Eqs.~\eqref{SoundKoebe} and \eqref{ShearKoebe} correspond to the longitudinal (energy diffusion) and, assuming Eq.~\eqref{PS-Trans}, the transverse (momentum diffusion) channels, respectively. The inequalities are fixed by pole-skipping and the radius of convergence. The lower bound in Eq.~\eqref{SoundKoebe} and the upper bound in Eq.~\eqref{ShearKoebe} have the form of the relation between $D$ and $v_B^2/\lambda_L$ first noticed by Blake \cite{Blake:2016wvh}. Moreover, our results imply that if a univalent diffusive $\omega(z)$ is entire (holomorphic everywhere except at infinity, so that $R\to\infty$), then $D = v_B^2/ \lambda_L$ identically. In terms of quasihydrodynamics \cite{Grozdanov:2018fic}, small $R$ is related to the relaxation time set by the leading gapped mode. Using Eq.~\eqref{C2Bound} for general $R$, we can now find bounds on $c_2$ (a third-order hydrodynamic coefficient \cite{Grozdanov:2015kqa,Diles:2019uft}):
\begin{equation}\label{C2Koebe}
0 \leq c_2 \leq \frac{D}{R}.
\end{equation}
The upper bound from either Eq.~\eqref{SoundKoebe} or Eq.~\eqref{ShearKoebe} eliminates $D$ from Eq.~\eqref{C2Koebe}. Simple algebraic manipulations give further bounds on $c_3$, $c_4$ and so on. If we can take $R \to \infty$, then $c_2 = 0$. Moreover, all $c_{n > 2}  = 0$ in this limit. Hence, for entire univalent $\omega_{\rm diff}(z)$, the dispersion relation truncates at the first order for all $\q^2$, with $D$ fixed by pole-skipping:
\begin{equation}\label{WExtreme}
\omega_{\rm diff} (\q^2) = - i D \q^2 = - i \frac{v_B^2}{\lambda_L} \q^2. 
\end{equation}

A theory that exhibits diffusive properties discussed here is a holographic model with broken translational invariance and energy diffusion \cite{Andrade:2013gsa}. At a special self-dual point in the parameter space of the background fields, symmetry enhancement allows us to analytically find the exact diffusive $\omega(z) = - i \pi T \left(1 - \sqrt{1-\frac{z}{\pi^2 T^2}} \right)$ \cite{Davison:2014lua}. Pole-skipping and hydrodynamic convergence in this theory were studied in \cite{Blake:2018leo,Grozdanov:2019uhi}, finding $z_0 = - 8\pi^2 T^2$, $v_B^2 = 1/2$, and $z_* =R = \pi^2 T^2$. The bounds implied by Eqs.~\eqref{SoundKoebe} and \eqref{C2Koebe}, along with the bounds on $c_3$, are then
\begin{align}
\frac{v_B^2}{\lambda_L} = \frac{1 }{4\pi T} &\leq D \leq  9 \frac{v_B^2}{\lambda_L}  = \frac{  9  }{4 \pi T }, \\
0 &\leq c_2 \leq \frac{D}{\pi^2 T^2} \leq \frac{9}{4 \pi^3 T^3}, \\
- \frac{27}{32 \pi^5 T^5} \leq - \frac{3 D}{8\pi^4 T^4} &\leq c_3 \leq \frac{D}{\pi^4 T^4} \leq \frac{9}{4 \pi^5 T^5}.
\end{align}
The actual values of $D = 1/2\pi T$, $c_2 = 1 / 8\pi^3 T^3$, and $c_3 = 1/16 \pi^5 T^5$ all satisfy the inequalities. 

{\bf Diffusion II: M\"{o}bius transformations.---}A general diffusive dispersion relation has multiple branch points and branch cuts. Generalising the scenario in which $U$ is determined by the group velocity conditions in Eqs.~\eqref{GroupVelocityConditionShear}--\eqref{GroupVelocityCondition}, let $U$ of $\omega_{\rm diff}(z)$ be a disk with a centre at $z=z_c$ and two boundary points at $z = z_c \pm z_b$ (on its closure), containing $z = 0$ and $z = z_0$, and with $z_c \in \mathbb{C}$ and $z_b \in \mathbb{R}_+$. $U$ can be mapped to $\mathbb{D}$ by the M\"{o}bius transformation $\zeta = \varphi(z)$, which we choose to be
\begin{equation}\label{MobiusDiffusion}
\varphi(z) = \frac{z_b z}{- z_c z + z_b^2 + z_c^2}, \quad \varphi^{-1}(\zeta) = \frac{\left(z_b^2 + z_c^2\right)\zeta}{z_b + z_c \zeta} ,
\end{equation}
satisfying $\varphi(0) = 0$ and mapping $z_c \pm i z_b \to \pm i$. We have $\partial^n_\zeta \varphi^{-1}(0) = n! (-z_c)^{n-1} (z_b^2  + z_c^2) / z_b^n$. All of the above bounds can now be easily constructed given specific $z_0$, $z_b$ and $z_c$. For example, Eq.~\eqref{DBoundGeneral} becomes
\begin{equation}
\frac{v_B^2}{\lambda_L} \left| 1 - \frac{z_c z_0}{z_b^2 + z_c^2}  \right| \CC_- \leq D \leq \frac{v_B^2}{\lambda_L} \left| 1 - \frac{z_c z_0}{z_b^2 + z_c^2}  \right| \CC_+ ,
\end{equation}
where $z_0 = \pm \lambda_L^2 / v_B^2$, depending on whether we use Eq.~\eqref{PS-Long} or Eq.~\eqref{PS-Trans}. $\CC_\pm$ are defined as 
\begin{equation}
\CC_\pm \equiv \left( 1 \pm |\zeta_0| \right)^2, \quad |\zeta_0| = \frac{\lambda_L^2}{v_B^2} \frac{z_b}{\left| - z_c z_0 + z_b^2 + z_c^2 \right|}.
\end{equation}

Of particular interest are cases with $z_c = 0$ so that $\varphi$ rescales a disk of radius $z_b = \min[|z_g| , R]$ to $\mathbb{D}$. The only non-zero $\partial^n_\zeta \varphi^{-1}(0)$ is then $\partial_\zeta \varphi^{-1}(0) = z_b$, and $b_n = z_b^{n-1} c_n / D$ for $n\geq 2$. The bounds on the coefficients of the series \eqref{WDiff} follow:
\begin{align}
\frac{v_B^2}{\lambda_L} \left( 1- \frac{1}{z_b} \frac{\lambda_L^2}{v_B^2} \right)^2 &\leq D \leq \frac{v_B^2}{\lambda_L} \left( 1+ \frac{1}{z_b} \frac{\lambda_L^2}{v_B^2} \right)^2 ,\label{DiffBounds2-1} \\
- \frac{nD}{z_b^{n-1}} &\leq c_{n\geq2} \leq  \frac{nD}{z_b^{n-1}} .\label{DiffBounds2-2}
\end{align} 
If the pole-skipping $z_0 \in U$, then by taking $z_b \to z_0$, we can at the very least establish that $0 \leq D \leq 4 v_B^2 / \lambda_L$. Also, as required, in the $z_b \to \infty$ limit, we again obtain the exact dispersion relation  \eqref{WExtreme}. If univalence of $f(\zeta)$ is ensured by $\re \,f'(\zeta) > 0$, then the bounds in Eqs.~\eqref{DiffBounds2-1} and \eqref{DiffBounds2-2} are improved:
\begin{align}
\frac{\lambda_L / z_b   }{\ln  \frac{e^{- \lambda_L^2 / z_b v_B^2} }{ \left(1 - \frac{\lambda_L^2}{z_b v_B^2}\right)^2 } }  &\leq D  \leq  \frac{\lambda_L/z_b  }{\ln  e^{- \lambda_L^2 / z_b v_B^2} \left(1 + \frac{\lambda_L^2}{z_b v_B^2}\right)^2  }   , \label{BoundsDiffusionReal1}\\
- \frac{2 D}{n z_b^{n-1}} &\leq c_{n\geq2} \leq  \frac{2 D}{n z_b^{n-1}}  . \label{BoundsDiffusionReal2}
\end{align} 
If $z_b \to \infty$, $\omega_{\rm diff} (\q^2)$ still reduces to the form in Eq.~\eqref{WExtreme}.

\begin{figure}[t!]
 \includegraphics[width=0.45\textwidth]{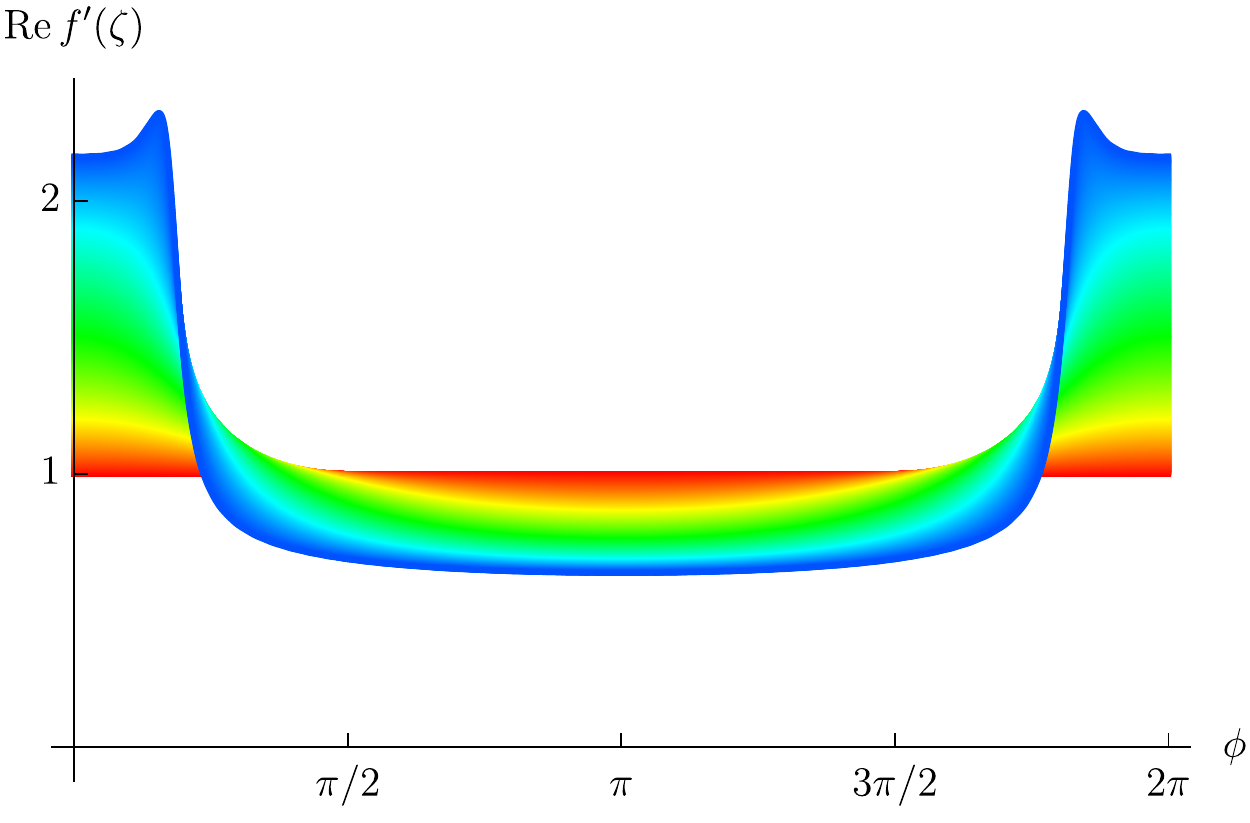}
\caption{
\label{fig:N4Diff}
The univalence condition $\re\,f'(\zeta)$, with $\zeta = |\zeta| e^{i\phi}$, plotted as a function of $\phi$ for momentum diffusion in ${\cal N}=4$ SYM theory. The colour gradient indicates different $|\zeta|$, from $|\zeta|=0$ (red) to $|\zeta| = 0.92$ (blue). We find that $\re\,f'(\zeta) > 0$ for all $|\zeta| < 1$, with $|\zeta| = 1$ mapped by $\varphi$ from $|z| = R$. }
\end{figure}

To demonstrate the existence of such theories, we consider momentum diffusion in two strongly coupled, large-$N$ theories at finite temperature: 3$d$ worldvolume theory of M2 branes and 4$d$ $\CN = 4$ supersymmetric Yang-Mills (SYM) theory. Diffusive $\omega_{\rm diff}(z)$ is determined by dual transverse metric fluctuations in 4$d$ \cite{Herzog:2002fn} and 5$d$ \cite{Policastro:2002se} Einstein-Hilbert theories with a negative cosmological constant and Anti-de Sitter-Schwarzschild black brane backgrounds. We check numerically that in both theories, $\re\,f'(z) > 0$ on their respective disks of hydrodynamic convergence, thereby establishing univalence for $|z| < z_b = R$. For the $\CN = 4$ SYM diffusion, we depict this in Fig. \ref{fig:N4Diff}.  The 3$d$ M2 brane case qualitatively matches the plot in Fig. \ref{fig:N4Diff}, with $R \approx 69.423 \, T^2$, $\lambda_L = 2\pi T$ and $v_B =\sqrt{3}/2 $.\footnote{\label{Foot2}Convergence of hydrodynamics in the holographic M2 brane theory is analysed by using the methods from Refs.~\cite{Grozdanov:2019kge,Grozdanov:2019uhi}. The transverse channel pole-skipping in Eq.~\eqref{PS-Trans} follows from the methods of \cite{Blake:2018leo,Grozdanov:2019uhi,Blake:2019otz}. We can prove analytically that $\omega_{\rm diff}(z) $ passes through an infinite sequence of pole-skipping points: $\omega_{\rm diff} (z_n) = - 2 \pi T i n$ at $z_n = \q_n^2 = 16 \pi^2 T^2 \sqrt{n} / 3 $ for all $n \in \mathbb{Z}_+ \cup \{0\}$.} In 4$d$ $\CN = 4$ SYM theory, $R \approx 87.800 \, T^2$, $\lambda_L = 2\pi T$ and $v_B = \sqrt{2/3}$ \cite{Grozdanov:2019kge,Grozdanov:2019uhi}. Given these values, we can numerically verify the validity of the bounds in Eqs.~\eqref{BoundsDiffusionReal1}--\eqref{BoundsDiffusionReal2}. For example, Eq.~\eqref{BoundsDiffusionReal1} evaluates to 
\begin{equation}
\frac{0.046}{T} \leq D  = \frac{1}{4\pi T} \approx \frac{0.080 }{ T} \leq \frac{0.201 }{ T}. 
\end{equation}
Moreover, the bounds become extremely tight as $n$ grows.  Assuming that the series coefficients $c_n$ become of the order of the bounds as $n\to\infty$ is consistent with the ratio test for convergence then giving $\lim_{n\to\infty} |c_n / c_{n+1}| = z_b$, which is the radius of convergence of Eq.~\eqref{WDiff}. 

\begin{figure}[t!]
 \includegraphics[width=0.45\textwidth]{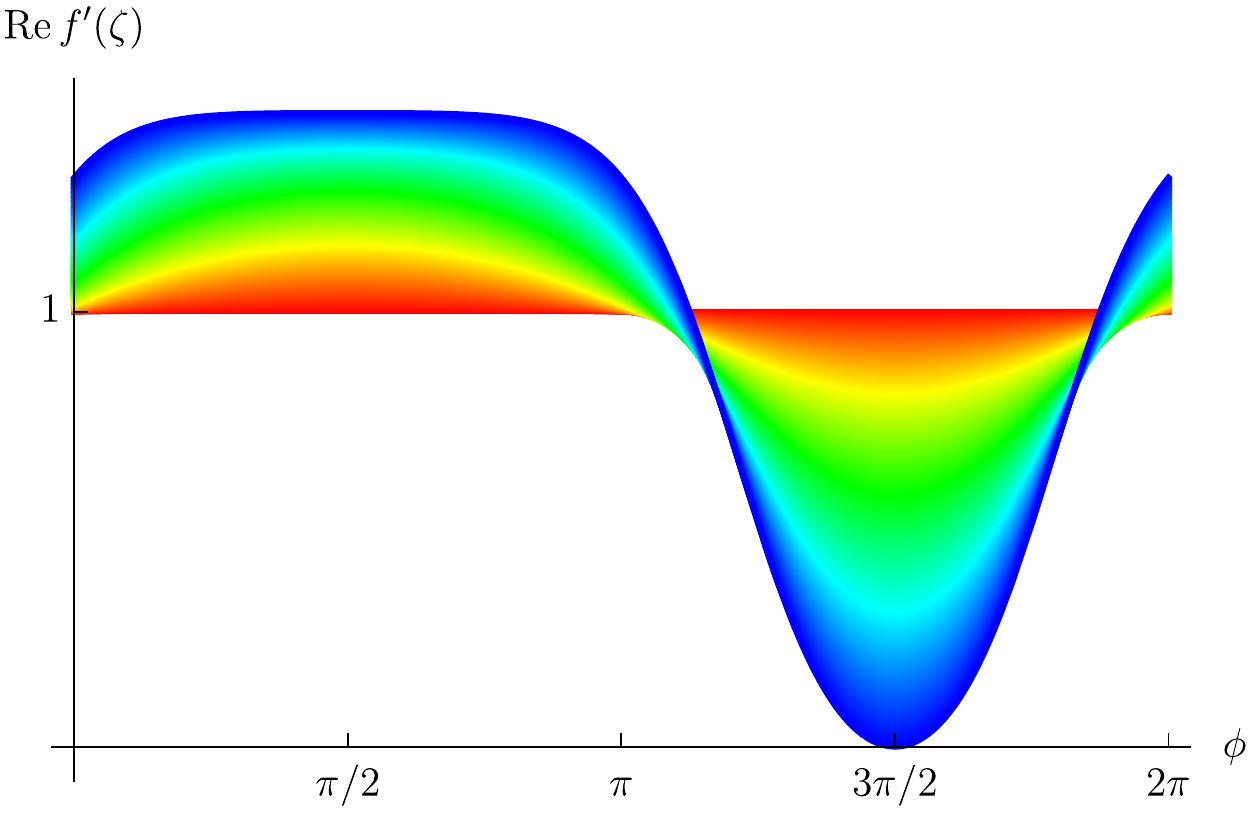}
\caption{
\label{fig:N4Sound}
$\re\,f'(\zeta)$, with $\zeta = |\zeta| e^{i\phi}$, plotted for sound in ${\cal N}=4$ SYM theory. The colour gradient runs from $|\zeta|=0$ (red) to $|\zeta| = 1$ (blue), with $|\zeta| = 1$ mapped by the $z_c = 0$ M\"{o}bius transformation $\varphi$ from $|z| = z_b = |z_g|$, where $v_g(z_g) = 0$.}
\end{figure}

{\bf Sound.---}By extending our holographic analysis to sound in the $\CN = 4$ SYM theory, we find that $\re\,f'(z) \ngtr 0$ on the hydrodynamic convergence disk $|z| < R$, where $R = 2 \sqrt{2} \pi T \approx 8.886 \, T$ \cite{Grozdanov:2019kge,Grozdanov:2019uhi}. Instead, $\re\,f'(z) > 0$ for $|z| < |z_g| < R$, with $z_g = q_g$ determined by the local condition in Eq.~\eqref{GroupVelocityCondition}. We depict the univalence condition in Fig. \ref{fig:N4Sound}. Numerically, we find that $z_g \approx - 3.791 \, i T$. Since $z_g$ lies within the hydrodynamic radius of convergence, its value can be crudely approximated by conformal first-order hydrodynamics: $z_g \approx - 3 i v_s / 4 D = - 5.441\, i T$ with $v_s = 1 / \sqrt{3}$ and $D = 1 / 4\pi T$. 

A crucial difference between this case and the diffusion above is that the pole-skipping $z_0 = i \lambda_L / v_B $ (cf. Eq.~\eqref{PS-Long}) is no longer in the $|z| < |z_g|$ disk of univalence $U$ (i.e., $|z_g| < |z_0| = \lambda_L / v_B \approx 7.695\, T$). However, it can be checked numerically that another univalent disk $z\in U$ can be chosen with $z_c \approx 2.548\, i T$ and $z_b \approx 6.338 \, T$ (cf. Eq.~\eqref{MobiusDiffusion}). The bounds on $\omega_{\rm sound}(z)$ then follow from Eqs.~\eqref{DBoundGeneral} and \eqref{deBranges} (not Eqs.~\eqref{DBoundGeneralReal} and \eqref{MacGregor2}, as $\re\,f'(\zeta) \ngtr 0$ for all $|\zeta| < 1$ after $\varphi: U \to \mathbb{D}$), with $|\omega_0| = \lambda_L$, as well as $\zeta_0$ and the derivatives of $\varphi^{-1}(0)$ computable from Eq.~\eqref{MobiusDiffusion}.

The maximally univalent sound analogue of Eq.~\eqref{WExtreme} is recovered when $z_c = 0 $ and $z_b \to \infty$. Then, we find an exact truncated dispersion relation $\omega_{\rm sound} (q) = \pm v_B q$. 

{\bf Bounds without pole-skipping.---}In the absence of pole-skipping considerations, we can derive bounds on transport purely in terms of the wave propagation speeds. For $U = \{z  \, | \, |z| < \min[|z_g| , R]\}$, with $z_g$ given by the group velocity conditions in Eqs.~\eqref{GroupVelocityConditionShear}--\eqref{GroupVelocityConditionSound} or in Eq.~\eqref{GroupVelocityCondition}, it follows that if the limit $|\zeta_0| \to 1$ exists, then Eq.~\eqref{DBoundGeneral} implies bounds expressed in terms of the phase velocities and momentum $\bar q$: $0 \leq D \leq 4 \left|  v_{ph}(\bar q^2) / \bar q  \right|$ and $0 \leq v_s \leq 4 \left| v_{ph}(\bar q )\right|$,  where $|\bar q | = \min[ |q_g|, |q_*| ]$. If we can use the inequalities from Eq.~\eqref{DBoundGeneralReal}, then $4$ in the upper bounds is improved to $1/(2\ln 2 - 1)$. Higher-order coefficients are bounded either by Eq.~\eqref{deBranges} or Eq.~\eqref{MacGregor2}. If $\bar q$ is the pole-skipping momentum $q_0$, we again recover the $z_b \to z_0$ limit of Eqs.~\eqref{DiffBounds2-1}--\eqref{BoundsDiffusionReal2}. 

For the final example, assume that there exists a class of theories that has the univalence properties of sound whereby $|\partial_\zeta \varphi^{-1}(0)| = 4 |\omega_0(z_0)| \sqrt{d-1} $, with $d$ as the number of spacetime dimensions. Moreover, assume that $\zeta_0 = \varphi(z_0) $ is infinitesimally close to the boundary of $\mathbb{D}$ and that the limit $|\zeta_0| \to 1$ again exists. Intriguingly, for theories satisfying these conditions, the growth theorem \eqref{DBoundGeneral} would then imply the following conformal upper bound on the speed of sound: $0 \leq v_s \leq \sqrt{1/(d-1)}$ \cite{Cherman:2009tw,Hohler:2009tv}, while also ensuring that the lower bound on $v_s$ is at $v_s = 0$. To better understand these simple univalence conditions and their powerful implications, it will be essential to find physical examples of theories (ideally, theories such as quantum chromodynamics \cite{Annala:2019puf}) that satisfy them or violate them \cite{Kulaxizi:2008jx,Hoyos:2016cob,Ecker:2017fyh,Ishii:2019gta}. Of particular interest should be any potential relation between the region $U$, the map $\varphi^{-1}(0)$ or $\omega_0(z_0)$, and the equation of state of the corresponding quantum field theory.

{\bf Discussion.---}To use the above construction of bounds, one must first establish univalence in $U$. Generically, as stated in Eqs.~\eqref{GroupVelocityConditionShear}--\eqref{GroupVelocityCondition}, hydrodynamic dispersion relations will be univalent up to at least the physically motivated group velocity conditions in complexified momentum space, which is sufficient to use the inequalities derived in this work. In holographic theories, this can be checked explicitly by numerical calculations. Finding more efficient methods for identifying (maximal or non-maximal) regions of univalence, possibly by directly using the associated bulk differential equations, remains an open problem. Another open problem is to explore univalence properties and emergent bounds in weakly coupled field theories and kinetic theory, as well as in quasihydrodynamic theories with long-lived gapped modes \cite{Grozdanov:2018fic}. It would also be interesting to understand whether univalence methods can be applied to non-linear and far-from-equilibrium hydrodynamic flows, as well as shed new light on the universality of the hydrodynamic attractors \cite{Heller:2015dha,Heller:2016rtz,Romatschke:2017vte,Denicol:2019lio}.

While pole-skipping was chosen in most examples due to our interest in relating bounds on transport to quantum chaos, as well as for convenience, any known value of $\omega_0 (z_0)$ in $U$ could also have been chosen. Two such examples were provided in the last section. Further simple examples can arise from the pole-skipping points without a clear connection to chaos. In fact, such choices may lead to more restrictive bounds. This naturally opens a general problem to find the tightest possible bounds within the scope of univalence techniques. As the univalence methods help pave the way towards more precise analytic explorations of transport, these and other questions will be addressed in the future.

{\bf Acknowledgements.---}I am grateful to Mike Blake, Richard Davison, Niko Jokela, Pavel Kovtun, Hong Liu, Andrei Starinets, Petar Tadi\'{c} and Aleksi Vuorinen for useful and stimulating discussions on related topics. This work was supported by U.S. DOE Grant No. DE-SC0011090 and the research programme P1-0402 of Slovenian Research Agency (ARRS).

{\bf Dedication.---}{\em In memory of my grandparents: \\
\v{Z}ivka and Ruse Grozdanov; Danica and Milan Rojko.}

\bibliographystyle{apsrev4-1}
\bibliography{Genbib}{}

\begin{thebibliography}{71}%
\makeatletter
\providecommand \@ifxundefined [1]{%
 \@ifx{#1\undefined}
}%
\providecommand \@ifnum [1]{%
 \ifnum #1\expandafter \@firstoftwo
 \else \expandafter \@secondoftwo
 \fi
}%
\providecommand \@ifx [1]{%
 \ifx #1\expandafter \@firstoftwo
 \else \expandafter \@secondoftwo
 \fi
}%
\providecommand \natexlab [1]{#1}%
\providecommand \enquote  [1]{``#1''}%
\providecommand \bibnamefont  [1]{#1}%
\providecommand \bibfnamefont [1]{#1}%
\providecommand \citenamefont [1]{#1}%
\providecommand \href@noop [0]{\@secondoftwo}%
\providecommand \href [0]{\begingroup \@sanitize@url \@href}%
\providecommand \@href[1]{\@@startlink{#1}\@@href}%
\providecommand \@@href[1]{\endgroup#1\@@endlink}%
\providecommand \@sanitize@url [0]{\catcode `\\12\catcode `\$12\catcode
  `\&12\catcode `\#12\catcode `\^12\catcode `\_12\catcode `\%12\relax}%
\providecommand \@@startlink[1]{}%
\providecommand \@@endlink[0]{}%
\providecommand \url  [0]{\begingroup\@sanitize@url \@url }%
\providecommand \@url [1]{\endgroup\@href {#1}{\urlprefix }}%
\providecommand \urlprefix  [0]{URL }%
\providecommand \Eprint [0]{\href }%
\providecommand \doibase [0]{http://dx.doi.org/}%
\providecommand \selectlanguage [0]{\@gobble}%
\providecommand \bibinfo  [0]{\@secondoftwo}%
\providecommand \bibfield  [0]{\@secondoftwo}%
\providecommand \translation [1]{[#1]}%
\providecommand \BibitemOpen [0]{}%
\providecommand \bibitemStop [0]{}%
\providecommand \bibitemNoStop [0]{.\EOS\space}%
\providecommand \EOS [0]{\spacefactor3000\relax}%
\providecommand \BibitemShut  [1]{\csname bibitem#1\endcsname}%
\let\auto@bib@innerbib\@empty
\bibitem [{\citenamefont {Sachdev}(2011)}]{sachdev_2011}%
  \BibitemOpen
  \bibfield  {author} {\bibinfo {author} {\bibfnamefont {S.}~\bibnamefont
  {Sachdev}},\ }\href {\doibase 10.1017/CBO9780511973765} {\emph {\bibinfo
  {title} {Quantum Phase Transitions}}},\ \bibinfo {edition} {2nd}\ ed.\
  (\bibinfo  {publisher} {Cambridge University Press},\ \bibinfo {year}
  {2011})\BibitemShut {NoStop}%
\bibitem [{\citenamefont {Ioffe}\ and\ \citenamefont
  {Regel}(1960)}]{ioffe1960non}%
  \BibitemOpen
  \bibfield  {author} {\bibinfo {author} {\bibfnamefont {A.}~\bibnamefont
  {Ioffe}}\ and\ \bibinfo {author} {\bibfnamefont {A.}~\bibnamefont {Regel}},\
  }\href@noop {} {\bibfield  {journal} {\bibinfo  {journal} {Prog. Semicond}\
  }\textbf {\bibinfo {volume} {4}},\ \bibinfo {pages} {237} (\bibinfo {year}
  {1960})}\BibitemShut {NoStop}%
\bibitem [{\citenamefont {Mott}(1972)}]{mott1972conduction}%
  \BibitemOpen
  \bibfield  {author} {\bibinfo {author} {\bibfnamefont {N.}~\bibnamefont
  {Mott}},\ }\href@noop {} {\bibfield  {journal} {\bibinfo  {journal}
  {Philosophical Magazine}\ }\textbf {\bibinfo {volume} {26}},\ \bibinfo
  {pages} {1015} (\bibinfo {year} {1972})}\BibitemShut {NoStop}%
\bibitem [{\citenamefont {Kovtun}\ \emph {et~al.}(2005)\citenamefont {Kovtun},
  \citenamefont {Son},\ and\ \citenamefont {Starinets}}]{Kovtun:2004de}%
  \BibitemOpen
  \bibfield  {author} {\bibinfo {author} {\bibfnamefont {P.}~\bibnamefont
  {Kovtun}}, \bibinfo {author} {\bibfnamefont {D.~T.}\ \bibnamefont {Son}}, \
  and\ \bibinfo {author} {\bibfnamefont {A.~O.}\ \bibnamefont {Starinets}},\
  }\href {\doibase 10.1103/PhysRevLett.94.111601} {\bibfield  {journal}
  {\bibinfo  {journal} {Phys. Rev. Lett.}\ }\textbf {\bibinfo {volume} {94}},\
  \bibinfo {pages} {111601} (\bibinfo {year} {2005})},\ \Eprint
  {http://arxiv.org/abs/hep-th/0405231} {arXiv:hep-th/0405231} \BibitemShut
  {NoStop}%
\bibitem [{\citenamefont {Hartnoll}(2015)}]{Hartnoll:2014lpa}%
  \BibitemOpen
  \bibfield  {author} {\bibinfo {author} {\bibfnamefont {S.~A.}\ \bibnamefont
  {Hartnoll}},\ }\href {\doibase 10.1038/nphys3174} {\bibfield  {journal}
  {\bibinfo  {journal} {Nature Phys.}\ }\textbf {\bibinfo {volume} {11}},\
  \bibinfo {pages} {54} (\bibinfo {year} {2015})},\ \Eprint
  {http://arxiv.org/abs/1405.3651} {arXiv:1405.3651 [cond-mat.str-el]}
  \BibitemShut {NoStop}%
\bibitem [{\citenamefont {Blake}(2016)}]{Blake:2016wvh}%
  \BibitemOpen
  \bibfield  {author} {\bibinfo {author} {\bibfnamefont {M.}~\bibnamefont
  {Blake}},\ }\href {\doibase 10.1103/PhysRevLett.117.091601} {\bibfield
  {journal} {\bibinfo  {journal} {Phys. Rev. Lett.}\ }\textbf {\bibinfo
  {volume} {117}},\ \bibinfo {pages} {091601} (\bibinfo {year} {2016})},\
  \Eprint {http://arxiv.org/abs/1603.08510} {arXiv:1603.08510 [hep-th]}
  \BibitemShut {NoStop}%
\bibitem [{\citenamefont {Zaanen}(2019)}]{Zaanen:2018edk}%
  \BibitemOpen
  \bibfield  {author} {\bibinfo {author} {\bibfnamefont {J.}~\bibnamefont
  {Zaanen}},\ }\href {\doibase 10.21468/SciPostPhys.6.5.061} {\bibfield
  {journal} {\bibinfo  {journal} {SciPost Phys.}\ }\textbf {\bibinfo {volume}
  {6}},\ \bibinfo {pages} {061} (\bibinfo {year} {2019})},\ \Eprint
  {http://arxiv.org/abs/1807.10951} {arXiv:1807.10951 [cond-mat.str-el]}
  \BibitemShut {NoStop}%
\bibitem [{\citenamefont {Trachenko}\ and\ \citenamefont
  {Brazhkin}(2020)}]{trachenko2020minimal}%
  \BibitemOpen
  \bibfield  {author} {\bibinfo {author} {\bibfnamefont {K.}~\bibnamefont
  {Trachenko}}\ and\ \bibinfo {author} {\bibfnamefont {V.}~\bibnamefont
  {Brazhkin}},\ }\href@noop {} {\bibfield  {journal} {\bibinfo  {journal}
  {Science Advances}\ }\textbf {\bibinfo {volume} {6}},\ \bibinfo {pages}
  {eaba3747} (\bibinfo {year} {2020})}\BibitemShut {NoStop}%
\bibitem [{\citenamefont {Baggioli}\ and\ \citenamefont
  {Li}(2020)}]{Baggioli:2020ljz}%
  \BibitemOpen
  \bibfield  {author} {\bibinfo {author} {\bibfnamefont {M.}~\bibnamefont
  {Baggioli}}\ and\ \bibinfo {author} {\bibfnamefont {W.-J.}\ \bibnamefont
  {Li}},\ }\href@noop {} {\  (\bibinfo {year} {2020})},\ \Eprint
  {http://arxiv.org/abs/2005.06482} {arXiv:2005.06482 [hep-th]} \BibitemShut
  {NoStop}%
\bibitem [{\citenamefont {Kovtun}\ \emph {et~al.}(2011)\citenamefont {Kovtun},
  \citenamefont {Moore},\ and\ \citenamefont {Romatschke}}]{Kovtun:2011np}%
  \BibitemOpen
  \bibfield  {author} {\bibinfo {author} {\bibfnamefont {P.}~\bibnamefont
  {Kovtun}}, \bibinfo {author} {\bibfnamefont {G.~D.}\ \bibnamefont {Moore}}, \
  and\ \bibinfo {author} {\bibfnamefont {P.}~\bibnamefont {Romatschke}},\
  }\href {\doibase 10.1103/PhysRevD.84.025006} {\bibfield  {journal} {\bibinfo
  {journal} {Phys. Rev. D}\ }\textbf {\bibinfo {volume} {84}},\ \bibinfo
  {pages} {025006} (\bibinfo {year} {2011})},\ \Eprint
  {http://arxiv.org/abs/1104.1586} {arXiv:1104.1586 [hep-ph]} \BibitemShut
  {NoStop}%
\bibitem [{\citenamefont {Chafin}\ and\ \citenamefont
  {Schaefer}(2013)}]{Chafin:2012eq}%
  \BibitemOpen
  \bibfield  {author} {\bibinfo {author} {\bibfnamefont {C.}~\bibnamefont
  {Chafin}}\ and\ \bibinfo {author} {\bibfnamefont {T.}~\bibnamefont
  {Schaefer}},\ }\href {\doibase 10.1103/PhysRevA.87.023629} {\bibfield
  {journal} {\bibinfo  {journal} {Phys. Rev. A}\ }\textbf {\bibinfo {volume}
  {87}},\ \bibinfo {pages} {023629} (\bibinfo {year} {2013})},\ \Eprint
  {http://arxiv.org/abs/1209.1006} {arXiv:1209.1006 [cond-mat.quant-gas]}
  \BibitemShut {NoStop}%
\bibitem [{\citenamefont {Kovtun}(2015)}]{Kovtun:2014nsa}%
  \BibitemOpen
  \bibfield  {author} {\bibinfo {author} {\bibfnamefont {P.}~\bibnamefont
  {Kovtun}},\ }\href {\doibase 10.1088/1751-8113/48/26/265002} {\bibfield
  {journal} {\bibinfo  {journal} {J. Phys. A}\ }\textbf {\bibinfo {volume}
  {48}},\ \bibinfo {pages} {265002} (\bibinfo {year} {2015})},\ \Eprint
  {http://arxiv.org/abs/1407.0690} {arXiv:1407.0690 [cond-mat.stat-mech]}
  \BibitemShut {NoStop}%
\bibitem [{\citenamefont {Martinez}\ and\ \citenamefont
  {Schaefer}(2017)}]{Martinez:2017jjf}%
  \BibitemOpen
  \bibfield  {author} {\bibinfo {author} {\bibfnamefont {M.}~\bibnamefont
  {Martinez}}\ and\ \bibinfo {author} {\bibfnamefont {T.}~\bibnamefont
  {Schaefer}},\ }\href {\doibase 10.1103/PhysRevA.96.063607} {\bibfield
  {journal} {\bibinfo  {journal} {Phys. Rev. A}\ }\textbf {\bibinfo {volume}
  {96}},\ \bibinfo {pages} {063607} (\bibinfo {year} {2017})},\ \Eprint
  {http://arxiv.org/abs/1708.01548} {arXiv:1708.01548 [cond-mat.quant-gas]}
  \BibitemShut {NoStop}%
\bibitem [{\citenamefont {Lucas}\ and\ \citenamefont
  {Steinberg}(2016)}]{Lucas:2016yfl}%
  \BibitemOpen
  \bibfield  {author} {\bibinfo {author} {\bibfnamefont {A.}~\bibnamefont
  {Lucas}}\ and\ \bibinfo {author} {\bibfnamefont {J.}~\bibnamefont
  {Steinberg}},\ }\href {\doibase 10.1007/JHEP10(2016)143} {\bibfield
  {journal} {\bibinfo  {journal} {JHEP}\ }\textbf {\bibinfo {volume} {10}},\
  \bibinfo {pages} {143} (\bibinfo {year} {2016})},\ \Eprint
  {http://arxiv.org/abs/1608.03286} {arXiv:1608.03286 [hep-th]} \BibitemShut
  {NoStop}%
\bibitem [{\citenamefont {Hartman}\ \emph {et~al.}(2017)\citenamefont
  {Hartman}, \citenamefont {Hartnoll},\ and\ \citenamefont
  {Mahajan}}]{Hartman:2017hhp}%
  \BibitemOpen
  \bibfield  {author} {\bibinfo {author} {\bibfnamefont {T.}~\bibnamefont
  {Hartman}}, \bibinfo {author} {\bibfnamefont {S.~A.}\ \bibnamefont
  {Hartnoll}}, \ and\ \bibinfo {author} {\bibfnamefont {R.}~\bibnamefont
  {Mahajan}},\ }\href {\doibase 10.1103/PhysRevLett.119.141601} {\bibfield
  {journal} {\bibinfo  {journal} {Phys. Rev. Lett.}\ }\textbf {\bibinfo
  {volume} {119}},\ \bibinfo {pages} {141601} (\bibinfo {year} {2017})},\
  \Eprint {http://arxiv.org/abs/1706.00019} {arXiv:1706.00019 [hep-th]}
  \BibitemShut {NoStop}%
\bibitem [{\citenamefont {Cherman}\ \emph {et~al.}(2009)\citenamefont
  {Cherman}, \citenamefont {Cohen},\ and\ \citenamefont
  {Nellore}}]{Cherman:2009tw}%
  \BibitemOpen
  \bibfield  {author} {\bibinfo {author} {\bibfnamefont {A.}~\bibnamefont
  {Cherman}}, \bibinfo {author} {\bibfnamefont {T.~D.}\ \bibnamefont {Cohen}},
  \ and\ \bibinfo {author} {\bibfnamefont {A.}~\bibnamefont {Nellore}},\ }\href
  {\doibase 10.1103/PhysRevD.80.066003} {\bibfield  {journal} {\bibinfo
  {journal} {Phys. Rev. D}\ }\textbf {\bibinfo {volume} {80}},\ \bibinfo
  {pages} {066003} (\bibinfo {year} {2009})},\ \Eprint
  {http://arxiv.org/abs/0905.0903} {arXiv:0905.0903 [hep-th]} \BibitemShut
  {NoStop}%
\bibitem [{\citenamefont {Hohler}\ and\ \citenamefont
  {Stephanov}(2009)}]{Hohler:2009tv}%
  \BibitemOpen
  \bibfield  {author} {\bibinfo {author} {\bibfnamefont {P.~M.}\ \bibnamefont
  {Hohler}}\ and\ \bibinfo {author} {\bibfnamefont {M.~A.}\ \bibnamefont
  {Stephanov}},\ }\href {\doibase 10.1103/PhysRevD.80.066002} {\bibfield
  {journal} {\bibinfo  {journal} {Phys. Rev. D}\ }\textbf {\bibinfo {volume}
  {80}},\ \bibinfo {pages} {066002} (\bibinfo {year} {2009})},\ \Eprint
  {http://arxiv.org/abs/0905.0900} {arXiv:0905.0900 [hep-th]} \BibitemShut
  {NoStop}%
\bibitem [{\citenamefont {Prosen}(2014)}]{Prosen_2014}%
  \BibitemOpen
  \bibfield  {author} {\bibinfo {author} {\bibfnamefont {T.}~\bibnamefont
  {Prosen}},\ }\href {\doibase 10.1103/physreve.89.012142} {\bibfield
  {journal} {\bibinfo  {journal} {Physical Review E}\ }\textbf {\bibinfo
  {volume} {89}} (\bibinfo {year} {2014}),\
  10.1103/physreve.89.012142}\BibitemShut {NoStop}%
\bibitem [{\citenamefont {Grozdanov}\ \emph {et~al.}(2015)\citenamefont
  {Grozdanov}, \citenamefont {Lucas}, \citenamefont {Sachdev},\ and\
  \citenamefont {Schalm}}]{Grozdanov:2015qia}%
  \BibitemOpen
  \bibfield  {author} {\bibinfo {author} {\bibfnamefont {S.}~\bibnamefont
  {Grozdanov}}, \bibinfo {author} {\bibfnamefont {A.}~\bibnamefont {Lucas}},
  \bibinfo {author} {\bibfnamefont {S.}~\bibnamefont {Sachdev}}, \ and\
  \bibinfo {author} {\bibfnamefont {K.}~\bibnamefont {Schalm}},\ }\href
  {\doibase 10.1103/PhysRevLett.115.221601} {\bibfield  {journal} {\bibinfo
  {journal} {Phys. Rev. Lett.}\ }\textbf {\bibinfo {volume} {115}},\ \bibinfo
  {pages} {221601} (\bibinfo {year} {2015})},\ \Eprint
  {http://arxiv.org/abs/1507.00003} {arXiv:1507.00003 [hep-th]} \BibitemShut
  {NoStop}%
\bibitem [{\citenamefont {Grozdanov}\ \emph {et~al.}(2016)\citenamefont
  {Grozdanov}, \citenamefont {Lucas},\ and\ \citenamefont
  {Schalm}}]{Grozdanov:2015djs}%
  \BibitemOpen
  \bibfield  {author} {\bibinfo {author} {\bibfnamefont {S.}~\bibnamefont
  {Grozdanov}}, \bibinfo {author} {\bibfnamefont {A.}~\bibnamefont {Lucas}}, \
  and\ \bibinfo {author} {\bibfnamefont {K.}~\bibnamefont {Schalm}},\ }\href
  {\doibase 10.1103/PhysRevD.93.061901} {\bibfield  {journal} {\bibinfo
  {journal} {Phys. Rev. D}\ }\textbf {\bibinfo {volume} {93}},\ \bibinfo
  {pages} {061901} (\bibinfo {year} {2016})},\ \Eprint
  {http://arxiv.org/abs/1511.05970} {arXiv:1511.05970 [hep-th]} \BibitemShut
  {NoStop}%
\bibitem [{\citenamefont {Maldacena}\ \emph {et~al.}(2016)\citenamefont
  {Maldacena}, \citenamefont {Shenker},\ and\ \citenamefont
  {Stanford}}]{Maldacena:2015waa}%
  \BibitemOpen
  \bibfield  {author} {\bibinfo {author} {\bibfnamefont {J.}~\bibnamefont
  {Maldacena}}, \bibinfo {author} {\bibfnamefont {S.~H.}\ \bibnamefont
  {Shenker}}, \ and\ \bibinfo {author} {\bibfnamefont {D.}~\bibnamefont
  {Stanford}},\ }\href {\doibase 10.1007/JHEP08(2016)106} {\bibfield  {journal}
  {\bibinfo  {journal} {JHEP}\ }\textbf {\bibinfo {volume} {08}},\ \bibinfo
  {pages} {106} (\bibinfo {year} {2016})},\ \Eprint
  {http://arxiv.org/abs/1503.01409} {arXiv:1503.01409 [hep-th]} \BibitemShut
  {NoStop}%
\bibitem [{\citenamefont {Kukuljan}\ \emph {et~al.}(2017)\citenamefont
  {Kukuljan}, \citenamefont {Grozdanov},\ and\ \citenamefont
  {Prosen}}]{Kukuljan:2017xag}%
  \BibitemOpen
  \bibfield  {author} {\bibinfo {author} {\bibfnamefont {I.}~\bibnamefont
  {Kukuljan}}, \bibinfo {author} {\bibfnamefont {S.}~\bibnamefont {Grozdanov}},
  \ and\ \bibinfo {author} {\bibfnamefont {T.}~\bibnamefont {Prosen}},\ }\href
  {\doibase 10.1103/PhysRevB.96.060301} {\bibfield  {journal} {\bibinfo
  {journal} {Phys. Rev.}\ }\textbf {\bibinfo {volume} {B96}},\ \bibinfo {pages}
  {060301} (\bibinfo {year} {2017})},\ \Eprint
  {http://arxiv.org/abs/1701.09147} {arXiv:1701.09147 [cond-mat.stat-mech]}
  \BibitemShut {NoStop}%
\bibitem [{\citenamefont {Duren}(2010)}]{duren2010univalent}%
  \BibitemOpen
  \bibfield  {author} {\bibinfo {author} {\bibfnamefont {P.}~\bibnamefont
  {Duren}},\ }\href {https://books.google.si/books?id=4EOlcQAACAAJ} {\emph
  {\bibinfo {title} {Univalent Functions}}},\ Grundlehren der mathematischen
  Wissenschaften\ (\bibinfo  {publisher} {Springer New York},\ \bibinfo {year}
  {2010})\BibitemShut {NoStop}%
\bibitem [{\citenamefont {Lehto}(2011)}]{lehto2011univalent}%
  \BibitemOpen
  \bibfield  {author} {\bibinfo {author} {\bibfnamefont {O.}~\bibnamefont
  {Lehto}},\ }\href {https://books.google.si/books?id=zsK7MAEACAAJ} {\emph
  {\bibinfo {title} {Univalent Functions and Teichm{\"u}ller Spaces}}},\
  Graduate Texts in Mathematics\ (\bibinfo  {publisher} {Springer New York},\
  \bibinfo {year} {2011})\BibitemShut {NoStop}%
\bibitem [{\citenamefont {Branges}(1985)}]{branges1985}%
  \BibitemOpen
  \bibfield  {author} {\bibinfo {author} {\bibfnamefont {L.}~\bibnamefont
  {Branges}},\ }\href {\doibase 10.1007/BF02392821} {\bibfield  {journal}
  {\bibinfo  {journal} {Acta Math.}\ }\textbf {\bibinfo {volume} {154}},\
  \bibinfo {pages} {137} (\bibinfo {year} {1985})}\BibitemShut {NoStop}%
\bibitem [{\citenamefont {Noshiro}(1934)}]{noshiro1934theory}%
  \BibitemOpen
  \bibfield  {author} {\bibinfo {author} {\bibfnamefont {K.}~\bibnamefont
  {Noshiro}},\ }\href@noop {} {\bibfield  {journal} {\bibinfo  {journal}
  {Journal of the Faculty of Science Hokkaido Imperial University. Ser. 1
  Mathematics}\ }\textbf {\bibinfo {volume} {2}},\ \bibinfo {pages} {129}
  (\bibinfo {year} {1934})}\BibitemShut {NoStop}%
\bibitem [{\citenamefont {Warschawski}(1935)}]{warschawski1935higher}%
  \BibitemOpen
  \bibfield  {author} {\bibinfo {author} {\bibfnamefont {S.~E.}\ \bibnamefont
  {Warschawski}},\ }\href@noop {} {\bibfield  {journal} {\bibinfo  {journal}
  {Transactions of the American Mathematical Society}\ }\textbf {\bibinfo
  {volume} {38}},\ \bibinfo {pages} {310} (\bibinfo {year} {1935})}\BibitemShut
  {NoStop}%
\bibitem [{\citenamefont {Macgregor}(1962)}]{macgregor1962functions}%
  \BibitemOpen
  \bibfield  {author} {\bibinfo {author} {\bibfnamefont {T.~H.}\ \bibnamefont
  {Macgregor}},\ }\href@noop {} {\bibfield  {journal} {\bibinfo  {journal}
  {Transactions of the American Mathematical Society}\ }\textbf {\bibinfo
  {volume} {104}},\ \bibinfo {pages} {532} (\bibinfo {year}
  {1962})}\BibitemShut {NoStop}%
\bibitem [{\citenamefont {Landau}\ and\ \citenamefont
  {Lifshits}(1987)}]{landau}%
  \BibitemOpen
  \bibfield  {author} {\bibinfo {author} {\bibfnamefont {L.}~\bibnamefont
  {Landau}}\ and\ \bibinfo {author} {\bibfnamefont {E.}~\bibnamefont
  {Lifshits}},\ }\href@noop {} {\emph {\bibinfo {title} {{Fluid Mechanics}}}}\
  (\bibinfo  {publisher} {Pergamon Press},\ \bibinfo {address} {New York},\
  \bibinfo {year} {1987})\BibitemShut {NoStop}%
\bibitem [{\citenamefont {Kovtun}(2012)}]{Kovtun:2012rj}%
  \BibitemOpen
  \bibfield  {author} {\bibinfo {author} {\bibfnamefont {P.}~\bibnamefont
  {Kovtun}},\ }\bibfield  {booktitle} {\emph {\bibinfo {booktitle} {{INT Summer
  School on Applications of String Theory Seattle, Washington, USA, July 18-29,
  2011}}},\ }\href {\doibase 10.1088/1751-8113/45/47/473001} {\bibfield
  {journal} {\bibinfo  {journal} {J. Phys.}\ }\textbf {\bibinfo {volume}
  {A45}},\ \bibinfo {pages} {473001} (\bibinfo {year} {2012})},\ \Eprint
  {http://arxiv.org/abs/1205.5040} {arXiv:1205.5040 [hep-th]} \BibitemShut
  {NoStop}%
\bibitem [{\citenamefont {Dubovsky}\ \emph {et~al.}(2012)\citenamefont
  {Dubovsky}, \citenamefont {Hui}, \citenamefont {Nicolis},\ and\ \citenamefont
  {Son}}]{Dubovsky:2011sj}%
  \BibitemOpen
  \bibfield  {author} {\bibinfo {author} {\bibfnamefont {S.}~\bibnamefont
  {Dubovsky}}, \bibinfo {author} {\bibfnamefont {L.}~\bibnamefont {Hui}},
  \bibinfo {author} {\bibfnamefont {A.}~\bibnamefont {Nicolis}}, \ and\
  \bibinfo {author} {\bibfnamefont {D.~T.}\ \bibnamefont {Son}},\ }\href
  {\doibase 10.1103/PhysRevD.85.085029} {\bibfield  {journal} {\bibinfo
  {journal} {Phys. Rev.}\ }\textbf {\bibinfo {volume} {D85}},\ \bibinfo {pages}
  {085029} (\bibinfo {year} {2012})},\ \Eprint {http://arxiv.org/abs/1107.0731}
  {arXiv:1107.0731 [hep-th]} \BibitemShut {NoStop}%
\bibitem [{\citenamefont {Grozdanov}\ and\ \citenamefont
  {Polonyi}(2015)}]{Grozdanov:2013dba}%
  \BibitemOpen
  \bibfield  {author} {\bibinfo {author} {\bibfnamefont {S.}~\bibnamefont
  {Grozdanov}}\ and\ \bibinfo {author} {\bibfnamefont {J.}~\bibnamefont
  {Polonyi}},\ }\href {\doibase 10.1103/PhysRevD.91.105031} {\bibfield
  {journal} {\bibinfo  {journal} {Phys. Rev.}\ }\textbf {\bibinfo {volume}
  {D91}},\ \bibinfo {pages} {105031} (\bibinfo {year} {2015})},\ \Eprint
  {http://arxiv.org/abs/1305.3670} {arXiv:1305.3670 [hep-th]} \BibitemShut
  {NoStop}%
\bibitem [{\citenamefont {Crossley}\ \emph {et~al.}(2017)\citenamefont
  {Crossley}, \citenamefont {Glorioso},\ and\ \citenamefont
  {Liu}}]{Crossley:2015evo}%
  \BibitemOpen
  \bibfield  {author} {\bibinfo {author} {\bibfnamefont {M.}~\bibnamefont
  {Crossley}}, \bibinfo {author} {\bibfnamefont {P.}~\bibnamefont {Glorioso}},
  \ and\ \bibinfo {author} {\bibfnamefont {H.}~\bibnamefont {Liu}},\ }\href
  {\doibase 10.1007/JHEP09(2017)095} {\bibfield  {journal} {\bibinfo  {journal}
  {JHEP}\ }\textbf {\bibinfo {volume} {09}},\ \bibinfo {pages} {095} (\bibinfo
  {year} {2017})},\ \Eprint {http://arxiv.org/abs/1511.03646} {arXiv:1511.03646
  [hep-th]} \BibitemShut {NoStop}%
\bibitem [{\citenamefont {Glorioso}\ \emph {et~al.}(2017)\citenamefont
  {Glorioso}, \citenamefont {Crossley},\ and\ \citenamefont
  {Liu}}]{Glorioso:2017fpd}%
  \BibitemOpen
  \bibfield  {author} {\bibinfo {author} {\bibfnamefont {P.}~\bibnamefont
  {Glorioso}}, \bibinfo {author} {\bibfnamefont {M.}~\bibnamefont {Crossley}},
  \ and\ \bibinfo {author} {\bibfnamefont {H.}~\bibnamefont {Liu}},\ }\href
  {\doibase 10.1007/JHEP09(2017)096} {\bibfield  {journal} {\bibinfo  {journal}
  {JHEP}\ }\textbf {\bibinfo {volume} {09}},\ \bibinfo {pages} {096} (\bibinfo
  {year} {2017})},\ \Eprint {http://arxiv.org/abs/1701.07817} {arXiv:1701.07817
  [hep-th]} \BibitemShut {NoStop}%
\bibitem [{\citenamefont {Haehl}\ \emph
  {et~al.}(2016{\natexlab{a}})\citenamefont {Haehl}, \citenamefont
  {Loganayagam},\ and\ \citenamefont {Rangamani}}]{Haehl:2015foa}%
  \BibitemOpen
  \bibfield  {author} {\bibinfo {author} {\bibfnamefont {F.~M.}\ \bibnamefont
  {Haehl}}, \bibinfo {author} {\bibfnamefont {R.}~\bibnamefont {Loganayagam}},
  \ and\ \bibinfo {author} {\bibfnamefont {M.}~\bibnamefont {Rangamani}},\
  }\href {\doibase 10.1007/JHEP01(2016)184} {\bibfield  {journal} {\bibinfo
  {journal} {JHEP}\ }\textbf {\bibinfo {volume} {01}},\ \bibinfo {pages} {184}
  (\bibinfo {year} {2016}{\natexlab{a}})},\ \Eprint
  {http://arxiv.org/abs/1510.02494} {arXiv:1510.02494 [hep-th]} \BibitemShut
  {NoStop}%
\bibitem [{\citenamefont {Haehl}\ \emph
  {et~al.}(2016{\natexlab{b}})\citenamefont {Haehl}, \citenamefont
  {Loganayagam},\ and\ \citenamefont {Rangamani}}]{Haehl:2015uoc}%
  \BibitemOpen
  \bibfield  {author} {\bibinfo {author} {\bibfnamefont {F.~M.}\ \bibnamefont
  {Haehl}}, \bibinfo {author} {\bibfnamefont {R.}~\bibnamefont {Loganayagam}},
  \ and\ \bibinfo {author} {\bibfnamefont {M.}~\bibnamefont {Rangamani}},\
  }\href {\doibase 10.1007/JHEP04(2016)039} {\bibfield  {journal} {\bibinfo
  {journal} {JHEP}\ }\textbf {\bibinfo {volume} {04}},\ \bibinfo {pages} {039}
  (\bibinfo {year} {2016}{\natexlab{b}})},\ \Eprint
  {http://arxiv.org/abs/1511.07809} {arXiv:1511.07809 [hep-th]} \BibitemShut
  {NoStop}%
\bibitem [{\citenamefont {Jensen}\ \emph {et~al.}(2018)\citenamefont {Jensen},
  \citenamefont {Pinzani-Fokeeva},\ and\ \citenamefont
  {Yarom}}]{Jensen:2017kzi}%
  \BibitemOpen
  \bibfield  {author} {\bibinfo {author} {\bibfnamefont {K.}~\bibnamefont
  {Jensen}}, \bibinfo {author} {\bibfnamefont {N.}~\bibnamefont
  {Pinzani-Fokeeva}}, \ and\ \bibinfo {author} {\bibfnamefont {A.}~\bibnamefont
  {Yarom}},\ }\href {\doibase 10.1007/JHEP09(2018)127} {\bibfield  {journal}
  {\bibinfo  {journal} {JHEP}\ }\textbf {\bibinfo {volume} {09}},\ \bibinfo
  {pages} {127} (\bibinfo {year} {2018})},\ \Eprint
  {http://arxiv.org/abs/1701.07436} {arXiv:1701.07436 [hep-th]} \BibitemShut
  {NoStop}%
\bibitem [{\citenamefont {Liu}\ and\ \citenamefont
  {Glorioso}(2018)}]{Glorioso:2018wxw}%
  \BibitemOpen
  \bibfield  {author} {\bibinfo {author} {\bibfnamefont {H.}~\bibnamefont
  {Liu}}\ and\ \bibinfo {author} {\bibfnamefont {P.}~\bibnamefont {Glorioso}},\
  }\bibfield  {booktitle} {\emph {\bibinfo {booktitle} {{Proceedings,
  Theoretical Advanced Study Institute in Elementary Particle Physics: Physics
  at the Fundamental Frontier (TASI 2017): Boulder, CO, USA, June 5-30,
  2017}}},\ }\href {\doibase 10.22323/1.305.0008} {\bibfield  {journal}
  {\bibinfo  {journal} {PoS}\ }\textbf {\bibinfo {volume} {TASI2017}},\
  \bibinfo {pages} {008} (\bibinfo {year} {2018})},\ \Eprint
  {http://arxiv.org/abs/1805.09331} {arXiv:1805.09331 [hep-th]} \BibitemShut
  {NoStop}%
\bibitem [{\citenamefont {Grozdanov}\ \emph {et~al.}(2017)\citenamefont
  {Grozdanov}, \citenamefont {Hofman},\ and\ \citenamefont
  {Iqbal}}]{Grozdanov:2016tdf}%
  \BibitemOpen
  \bibfield  {author} {\bibinfo {author} {\bibfnamefont {S.}~\bibnamefont
  {Grozdanov}}, \bibinfo {author} {\bibfnamefont {D.~M.}\ \bibnamefont
  {Hofman}}, \ and\ \bibinfo {author} {\bibfnamefont {N.}~\bibnamefont
  {Iqbal}},\ }\href {\doibase 10.1103/PhysRevD.95.096003} {\bibfield  {journal}
  {\bibinfo  {journal} {Phys. Rev. D}\ }\textbf {\bibinfo {volume} {95}},\
  \bibinfo {pages} {096003} (\bibinfo {year} {2017})},\ \Eprint
  {http://arxiv.org/abs/1610.07392} {arXiv:1610.07392 [hep-th]} \BibitemShut
  {NoStop}%
\bibitem [{\citenamefont {Chen-Lin}\ \emph {et~al.}(2019)\citenamefont
  {Chen-Lin}, \citenamefont {Delacretaz},\ and\ \citenamefont
  {Hartnoll}}]{Chen-Lin:2018kfl}%
  \BibitemOpen
  \bibfield  {author} {\bibinfo {author} {\bibfnamefont {X.}~\bibnamefont
  {Chen-Lin}}, \bibinfo {author} {\bibfnamefont {L.~V.}\ \bibnamefont
  {Delacretaz}}, \ and\ \bibinfo {author} {\bibfnamefont {S.~A.}\ \bibnamefont
  {Hartnoll}},\ }\href {\doibase 10.1103/PhysRevLett.122.091602} {\bibfield
  {journal} {\bibinfo  {journal} {Phys. Rev. Lett.}\ }\textbf {\bibinfo
  {volume} {122}},\ \bibinfo {pages} {091602} (\bibinfo {year} {2019})},\
  \Eprint {http://arxiv.org/abs/1811.12540} {arXiv:1811.12540 [hep-th]}
  \BibitemShut {NoStop}%
\bibitem [{\citenamefont {Kovtun}\ and\ \citenamefont
  {Yaffe}(2003)}]{Kovtun:2003vj}%
  \BibitemOpen
  \bibfield  {author} {\bibinfo {author} {\bibfnamefont {P.}~\bibnamefont
  {Kovtun}}\ and\ \bibinfo {author} {\bibfnamefont {L.~G.}\ \bibnamefont
  {Yaffe}},\ }\href {\doibase 10.1103/PhysRevD.68.025007} {\bibfield  {journal}
  {\bibinfo  {journal} {Phys. Rev.}\ }\textbf {\bibinfo {volume} {D68}},\
  \bibinfo {pages} {025007} (\bibinfo {year} {2003})},\ \Eprint
  {http://arxiv.org/abs/hep-th/0303010} {arXiv:hep-th/0303010 [hep-th]}
  \BibitemShut {NoStop}%
\bibitem [{\citenamefont {Delacretaz}(2020)}]{Delacretaz:2020nit}%
  \BibitemOpen
  \bibfield  {author} {\bibinfo {author} {\bibfnamefont {L.~V.}\ \bibnamefont
  {Delacretaz}},\ }\href@noop {} {\  (\bibinfo {year} {2020})},\ \Eprint
  {http://arxiv.org/abs/2006.01139} {arXiv:2006.01139 [hep-th]} \BibitemShut
  {NoStop}%
\bibitem [{\citenamefont {Grozdanov}\ \emph
  {et~al.}(2019{\natexlab{a}})\citenamefont {Grozdanov}, \citenamefont
  {Kovtun}, \citenamefont {Starinets},\ and\ \citenamefont
  {Tadi\'{c}}}]{Grozdanov:2019kge}%
  \BibitemOpen
  \bibfield  {author} {\bibinfo {author} {\bibfnamefont {S.}~\bibnamefont
  {Grozdanov}}, \bibinfo {author} {\bibfnamefont {P.~K.}\ \bibnamefont
  {Kovtun}}, \bibinfo {author} {\bibfnamefont {A.~O.}\ \bibnamefont
  {Starinets}}, \ and\ \bibinfo {author} {\bibfnamefont {P.}~\bibnamefont
  {Tadi\'{c}}},\ }\href {\doibase 10.1103/PhysRevLett.122.251601} {\bibfield
  {journal} {\bibinfo  {journal} {Phys. Rev. Lett.}\ }\textbf {\bibinfo
  {volume} {122}},\ \bibinfo {pages} {251601} (\bibinfo {year}
  {2019}{\natexlab{a}})},\ \Eprint {http://arxiv.org/abs/1904.01018}
  {arXiv:1904.01018 [hep-th]} \BibitemShut {NoStop}%
\bibitem [{\citenamefont {Grozdanov}\ \emph
  {et~al.}(2019{\natexlab{b}})\citenamefont {Grozdanov}, \citenamefont
  {Kovtun}, \citenamefont {Starinets},\ and\ \citenamefont
  {Tadi\'c}}]{Grozdanov:2019uhi}%
  \BibitemOpen
  \bibfield  {author} {\bibinfo {author} {\bibfnamefont {S.}~\bibnamefont
  {Grozdanov}}, \bibinfo {author} {\bibfnamefont {P.~K.}\ \bibnamefont
  {Kovtun}}, \bibinfo {author} {\bibfnamefont {A.~O.}\ \bibnamefont
  {Starinets}}, \ and\ \bibinfo {author} {\bibfnamefont {P.}~\bibnamefont
  {Tadi\'c}},\ }\href {\doibase 10.1007/JHEP11(2019)097} {\bibfield  {journal}
  {\bibinfo  {journal} {JHEP}\ }\textbf {\bibinfo {volume} {11}},\ \bibinfo
  {pages} {097} (\bibinfo {year} {2019}{\natexlab{b}})},\ \Eprint
  {http://arxiv.org/abs/1904.12862} {arXiv:1904.12862 [hep-th]} \BibitemShut
  {NoStop}%
\bibitem [{\citenamefont {Withers}(2018)}]{Withers:2018srf}%
  \BibitemOpen
  \bibfield  {author} {\bibinfo {author} {\bibfnamefont {B.}~\bibnamefont
  {Withers}},\ }\href {\doibase 10.1007/JHEP06(2018)059} {\bibfield  {journal}
  {\bibinfo  {journal} {JHEP}\ }\textbf {\bibinfo {volume} {06}},\ \bibinfo
  {pages} {059} (\bibinfo {year} {2018})},\ \Eprint
  {http://arxiv.org/abs/1803.08058} {arXiv:1803.08058 [hep-th]} \BibitemShut
  {NoStop}%
\bibitem [{\citenamefont {Heller}\ \emph {et~al.}(2020)\citenamefont {Heller},
  \citenamefont {Serantes}, \citenamefont {Spali\'nski}, \citenamefont
  {Svensson},\ and\ \citenamefont {Withers}}]{Heller:2020uuy}%
  \BibitemOpen
  \bibfield  {author} {\bibinfo {author} {\bibfnamefont {M.~P.}\ \bibnamefont
  {Heller}}, \bibinfo {author} {\bibfnamefont {A.}~\bibnamefont {Serantes}},
  \bibinfo {author} {\bibfnamefont {M.}~\bibnamefont {Spali\'nski}}, \bibinfo
  {author} {\bibfnamefont {V.}~\bibnamefont {Svensson}}, \ and\ \bibinfo
  {author} {\bibfnamefont {B.}~\bibnamefont {Withers}},\ }\href@noop {} {\
  (\bibinfo {year} {2020})},\ \Eprint {http://arxiv.org/abs/2007.05524}
  {arXiv:2007.05524 [hep-th]} \BibitemShut {NoStop}%
\bibitem [{\citenamefont {Abbasi}\ and\ \citenamefont
  {Tahery}(2020)}]{Abbasi:2020ykq}%
  \BibitemOpen
  \bibfield  {author} {\bibinfo {author} {\bibfnamefont {N.}~\bibnamefont
  {Abbasi}}\ and\ \bibinfo {author} {\bibfnamefont {S.}~\bibnamefont
  {Tahery}},\ }\href@noop {} {\  (\bibinfo {year} {2020})},\ \Eprint
  {http://arxiv.org/abs/2007.10024} {arXiv:2007.10024 [hep-th]} \BibitemShut
  {NoStop}%
\bibitem [{\citenamefont {Jansen}\ and\ \citenamefont
  {Pantelidou}(2020)}]{1809177}%
  \BibitemOpen
  \bibfield  {author} {\bibinfo {author} {\bibfnamefont {A.}~\bibnamefont
  {Jansen}}\ and\ \bibinfo {author} {\bibfnamefont {C.}~\bibnamefont
  {Pantelidou}},\ }\href@noop {} {\  (\bibinfo {year} {2020})},\ \Eprint
  {http://arxiv.org/abs/2007.14418} {arXiv:2007.14418 [hep-th]} \BibitemShut
  {NoStop}%
\bibitem [{\citenamefont {Krotscheck}\ and\ \citenamefont
  {Kundt}(1978)}]{krotscheck1978causality}%
  \BibitemOpen
  \bibfield  {author} {\bibinfo {author} {\bibfnamefont {E.}~\bibnamefont
  {Krotscheck}}\ and\ \bibinfo {author} {\bibfnamefont {W.}~\bibnamefont
  {Kundt}},\ }\href@noop {} {\bibfield  {journal} {\bibinfo  {journal}
  {Communications in Mathematical Physics}\ }\textbf {\bibinfo {volume} {60}},\
  \bibinfo {pages} {171} (\bibinfo {year} {1978})}\BibitemShut {NoStop}%
\bibitem [{\citenamefont {Grozdanov}\ \emph {et~al.}(2018)\citenamefont
  {Grozdanov}, \citenamefont {Schalm},\ and\ \citenamefont
  {Scopelliti}}]{Grozdanov:2017ajz}%
  \BibitemOpen
  \bibfield  {author} {\bibinfo {author} {\bibfnamefont {S.}~\bibnamefont
  {Grozdanov}}, \bibinfo {author} {\bibfnamefont {K.}~\bibnamefont {Schalm}}, \
  and\ \bibinfo {author} {\bibfnamefont {V.}~\bibnamefont {Scopelliti}},\
  }\href {\doibase 10.1103/PhysRevLett.120.231601} {\bibfield  {journal}
  {\bibinfo  {journal} {Phys. Rev. Lett.}\ }\textbf {\bibinfo {volume} {120}},\
  \bibinfo {pages} {231601} (\bibinfo {year} {2018})},\ \Eprint
  {http://arxiv.org/abs/1710.00921} {arXiv:1710.00921 [hep-th]} \BibitemShut
  {NoStop}%
\bibitem [{\citenamefont {Blake}\ \emph
  {et~al.}(2018{\natexlab{a}})\citenamefont {Blake}, \citenamefont {Lee},\ and\
  \citenamefont {Liu}}]{Blake:2017ris}%
  \BibitemOpen
  \bibfield  {author} {\bibinfo {author} {\bibfnamefont {M.}~\bibnamefont
  {Blake}}, \bibinfo {author} {\bibfnamefont {H.}~\bibnamefont {Lee}}, \ and\
  \bibinfo {author} {\bibfnamefont {H.}~\bibnamefont {Liu}},\ }\href {\doibase
  10.1007/JHEP10(2018)127} {\bibfield  {journal} {\bibinfo  {journal} {JHEP}\
  }\textbf {\bibinfo {volume} {10}},\ \bibinfo {pages} {127} (\bibinfo {year}
  {2018}{\natexlab{a}})},\ \Eprint {http://arxiv.org/abs/1801.00010}
  {arXiv:1801.00010 [hep-th]} \BibitemShut {NoStop}%
\bibitem [{\citenamefont {Blake}\ \emph
  {et~al.}(2018{\natexlab{b}})\citenamefont {Blake}, \citenamefont {Davison},
  \citenamefont {Grozdanov},\ and\ \citenamefont {Liu}}]{Blake:2018leo}%
  \BibitemOpen
  \bibfield  {author} {\bibinfo {author} {\bibfnamefont {M.}~\bibnamefont
  {Blake}}, \bibinfo {author} {\bibfnamefont {R.~A.}\ \bibnamefont {Davison}},
  \bibinfo {author} {\bibfnamefont {S.}~\bibnamefont {Grozdanov}}, \ and\
  \bibinfo {author} {\bibfnamefont {H.}~\bibnamefont {Liu}},\ }\href {\doibase
  10.1007/JHEP10(2018)035} {\bibfield  {journal} {\bibinfo  {journal} {JHEP}\
  }\textbf {\bibinfo {volume} {10}},\ \bibinfo {pages} {035} (\bibinfo {year}
  {2018}{\natexlab{b}})},\ \Eprint {http://arxiv.org/abs/1809.01169}
  {arXiv:1809.01169 [hep-th]} \BibitemShut {NoStop}%
\bibitem [{\citenamefont {Grozdanov}(2019)}]{Grozdanov:2018kkt}%
  \BibitemOpen
  \bibfield  {author} {\bibinfo {author} {\bibfnamefont {S.}~\bibnamefont
  {Grozdanov}},\ }\href {\doibase 10.1007/JHEP01(2019)048} {\bibfield
  {journal} {\bibinfo  {journal} {JHEP}\ }\textbf {\bibinfo {volume} {01}},\
  \bibinfo {pages} {048} (\bibinfo {year} {2019})},\ \Eprint
  {http://arxiv.org/abs/1811.09641} {arXiv:1811.09641 [hep-th]} \BibitemShut
  {NoStop}%
\bibitem [{\citenamefont {Shenker}\ and\ \citenamefont
  {Stanford}(2014)}]{Shenker:2013pqa}%
  \BibitemOpen
  \bibfield  {author} {\bibinfo {author} {\bibfnamefont {S.~H.}\ \bibnamefont
  {Shenker}}\ and\ \bibinfo {author} {\bibfnamefont {D.}~\bibnamefont
  {Stanford}},\ }\href {\doibase 10.1007/JHEP03(2014)067} {\bibfield  {journal}
  {\bibinfo  {journal} {JHEP}\ }\textbf {\bibinfo {volume} {03}},\ \bibinfo
  {pages} {067} (\bibinfo {year} {2014})},\ \Eprint
  {http://arxiv.org/abs/1306.0622} {arXiv:1306.0622 [hep-th]} \BibitemShut
  {NoStop}%
\bibitem [{\citenamefont {Blake}\ \emph {et~al.}(2020)\citenamefont {Blake},
  \citenamefont {Davison},\ and\ \citenamefont {Vegh}}]{Blake:2019otz}%
  \BibitemOpen
  \bibfield  {author} {\bibinfo {author} {\bibfnamefont {M.}~\bibnamefont
  {Blake}}, \bibinfo {author} {\bibfnamefont {R.~A.}\ \bibnamefont {Davison}},
  \ and\ \bibinfo {author} {\bibfnamefont {D.}~\bibnamefont {Vegh}},\ }\href
  {\doibase 10.1007/JHEP01(2020)077} {\bibfield  {journal} {\bibinfo  {journal}
  {JHEP}\ }\textbf {\bibinfo {volume} {01}},\ \bibinfo {pages} {077} (\bibinfo
  {year} {2020})},\ \Eprint {http://arxiv.org/abs/1904.12883} {arXiv:1904.12883
  [hep-th]} \BibitemShut {NoStop}%
\bibitem [{\citenamefont {Grozdanov}\ \emph
  {et~al.}(2019{\natexlab{c}})\citenamefont {Grozdanov}, \citenamefont
  {Lucas},\ and\ \citenamefont {Poovuttikul}}]{Grozdanov:2018fic}%
  \BibitemOpen
  \bibfield  {author} {\bibinfo {author} {\bibfnamefont {S.}~\bibnamefont
  {Grozdanov}}, \bibinfo {author} {\bibfnamefont {A.}~\bibnamefont {Lucas}}, \
  and\ \bibinfo {author} {\bibfnamefont {N.}~\bibnamefont {Poovuttikul}},\
  }\href {\doibase 10.1103/PhysRevD.99.086012} {\bibfield  {journal} {\bibinfo
  {journal} {Phys. Rev. D}\ }\textbf {\bibinfo {volume} {99}},\ \bibinfo
  {pages} {086012} (\bibinfo {year} {2019}{\natexlab{c}})},\ \Eprint
  {http://arxiv.org/abs/1810.10016} {arXiv:1810.10016 [hep-th]} \BibitemShut
  {NoStop}%
\bibitem [{\citenamefont {Grozdanov}\ and\ \citenamefont
  {Kaplis}(2016)}]{Grozdanov:2015kqa}%
  \BibitemOpen
  \bibfield  {author} {\bibinfo {author} {\bibfnamefont {S.}~\bibnamefont
  {Grozdanov}}\ and\ \bibinfo {author} {\bibfnamefont {N.}~\bibnamefont
  {Kaplis}},\ }\href {\doibase 10.1103/PhysRevD.93.066012} {\bibfield
  {journal} {\bibinfo  {journal} {Phys. Rev.}\ }\textbf {\bibinfo {volume}
  {D93}},\ \bibinfo {pages} {066012} (\bibinfo {year} {2016})},\ \Eprint
  {http://arxiv.org/abs/1507.02461} {arXiv:1507.02461 [hep-th]} \BibitemShut
  {NoStop}%
\bibitem [{\citenamefont {Diles}\ \emph {et~al.}(2020)\citenamefont {Diles},
  \citenamefont {Mamani}, \citenamefont {Miranda},\ and\ \citenamefont
  {Zanchin}}]{Diles:2019uft}%
  \BibitemOpen
  \bibfield  {author} {\bibinfo {author} {\bibfnamefont {S.~M.}\ \bibnamefont
  {Diles}}, \bibinfo {author} {\bibfnamefont {L.~A.}\ \bibnamefont {Mamani}},
  \bibinfo {author} {\bibfnamefont {A.~S.}\ \bibnamefont {Miranda}}, \ and\
  \bibinfo {author} {\bibfnamefont {V.~T.}\ \bibnamefont {Zanchin}},\ }\href
  {\doibase 10.1007/JHEP05(2020)019} {\bibfield  {journal} {\bibinfo  {journal}
  {JHEP}\ }\textbf {\bibinfo {volume} {05}},\ \bibinfo {pages} {019} (\bibinfo
  {year} {2020})},\ \Eprint {http://arxiv.org/abs/1909.05199} {arXiv:1909.05199
  [hep-th]} \BibitemShut {NoStop}%
\bibitem [{\citenamefont {Andrade}\ and\ \citenamefont
  {Withers}(2014)}]{Andrade:2013gsa}%
  \BibitemOpen
  \bibfield  {author} {\bibinfo {author} {\bibfnamefont {T.}~\bibnamefont
  {Andrade}}\ and\ \bibinfo {author} {\bibfnamefont {B.}~\bibnamefont
  {Withers}},\ }\href {\doibase 10.1007/JHEP05(2014)101} {\bibfield  {journal}
  {\bibinfo  {journal} {JHEP}\ }\textbf {\bibinfo {volume} {05}},\ \bibinfo
  {pages} {101} (\bibinfo {year} {2014})},\ \Eprint
  {http://arxiv.org/abs/1311.5157} {arXiv:1311.5157 [hep-th]} \BibitemShut
  {NoStop}%
\bibitem [{\citenamefont {Davison}\ and\ \citenamefont
  {Gouteraux}(2015)}]{Davison:2014lua}%
  \BibitemOpen
  \bibfield  {author} {\bibinfo {author} {\bibfnamefont {R.~A.}\ \bibnamefont
  {Davison}}\ and\ \bibinfo {author} {\bibfnamefont {B.}~\bibnamefont
  {Gouteraux}},\ }\href {\doibase 10.1007/JHEP01(2015)039} {\bibfield
  {journal} {\bibinfo  {journal} {JHEP}\ }\textbf {\bibinfo {volume} {01}},\
  \bibinfo {pages} {039} (\bibinfo {year} {2015})},\ \Eprint
  {http://arxiv.org/abs/1411.1062} {arXiv:1411.1062 [hep-th]} \BibitemShut
  {NoStop}%
\bibitem [{\citenamefont {Herzog}(2002)}]{Herzog:2002fn}%
  \BibitemOpen
  \bibfield  {author} {\bibinfo {author} {\bibfnamefont {C.~P.}\ \bibnamefont
  {Herzog}},\ }\href {\doibase 10.1088/1126-6708/2002/12/026} {\bibfield
  {journal} {\bibinfo  {journal} {JHEP}\ }\textbf {\bibinfo {volume} {12}},\
  \bibinfo {pages} {026} (\bibinfo {year} {2002})},\ \Eprint
  {http://arxiv.org/abs/hep-th/0210126} {arXiv:hep-th/0210126} \BibitemShut
  {NoStop}%
\bibitem [{\citenamefont {Policastro}\ \emph {et~al.}(2002)\citenamefont
  {Policastro}, \citenamefont {Son},\ and\ \citenamefont
  {Starinets}}]{Policastro:2002se}%
  \BibitemOpen
  \bibfield  {author} {\bibinfo {author} {\bibfnamefont {G.}~\bibnamefont
  {Policastro}}, \bibinfo {author} {\bibfnamefont {D.~T.}\ \bibnamefont {Son}},
  \ and\ \bibinfo {author} {\bibfnamefont {A.~O.}\ \bibnamefont {Starinets}},\
  }\href {\doibase 10.1088/1126-6708/2002/09/043} {\bibfield  {journal}
  {\bibinfo  {journal} {JHEP}\ }\textbf {\bibinfo {volume} {09}},\ \bibinfo
  {pages} {043} (\bibinfo {year} {2002})},\ \Eprint
  {http://arxiv.org/abs/hep-th/0205052} {arXiv:hep-th/0205052} \BibitemShut
  {NoStop}%
\bibitem [{\citenamefont {Annala}\ \emph {et~al.}(2020)\citenamefont {Annala},
  \citenamefont {Gorda}, \citenamefont {Kurkela}, \citenamefont
  {N{\"a}ttil{\"a}},\ and\ \citenamefont {Vuorinen}}]{Annala:2019puf}%
  \BibitemOpen
  \bibfield  {author} {\bibinfo {author} {\bibfnamefont {E.}~\bibnamefont
  {Annala}}, \bibinfo {author} {\bibfnamefont {T.}~\bibnamefont {Gorda}},
  \bibinfo {author} {\bibfnamefont {A.}~\bibnamefont {Kurkela}}, \bibinfo
  {author} {\bibfnamefont {J.}~\bibnamefont {N{\"a}ttil{\"a}}}, \ and\ \bibinfo
  {author} {\bibfnamefont {A.}~\bibnamefont {Vuorinen}},\ }\href {\doibase
  10.1038/s41567-020-0914-9} {\bibfield  {journal} {\bibinfo  {journal} {Nature
  Phys.}\ } (\bibinfo {year} {2020}),\ 10.1038/s41567-020-0914-9},\ \Eprint
  {http://arxiv.org/abs/1903.09121} {arXiv:1903.09121 [astro-ph.HE]}
  \BibitemShut {NoStop}%
\bibitem [{\citenamefont {Kulaxizi}\ and\ \citenamefont
  {Parnachev}(2009)}]{Kulaxizi:2008jx}%
  \BibitemOpen
  \bibfield  {author} {\bibinfo {author} {\bibfnamefont {M.}~\bibnamefont
  {Kulaxizi}}\ and\ \bibinfo {author} {\bibfnamefont {A.}~\bibnamefont
  {Parnachev}},\ }\href {\doibase 10.1016/j.nuclphysb.2009.02.016} {\bibfield
  {journal} {\bibinfo  {journal} {Nucl. Phys. B}\ }\textbf {\bibinfo {volume}
  {815}},\ \bibinfo {pages} {125} (\bibinfo {year} {2009})},\ \Eprint
  {http://arxiv.org/abs/0811.2262} {arXiv:0811.2262 [hep-th]} \BibitemShut
  {NoStop}%
\bibitem [{\citenamefont {Hoyos}\ \emph {et~al.}(2016)\citenamefont {Hoyos},
  \citenamefont {Jokela}, \citenamefont {Rodr\'{i}guez~Fern\'{a}ndez},\ and\
  \citenamefont {Vuorinen}}]{Hoyos:2016cob}%
  \BibitemOpen
  \bibfield  {author} {\bibinfo {author} {\bibfnamefont {C.}~\bibnamefont
  {Hoyos}}, \bibinfo {author} {\bibfnamefont {N.}~\bibnamefont {Jokela}},
  \bibinfo {author} {\bibfnamefont {D.}~\bibnamefont
  {Rodr\'{i}guez~Fern\'{a}ndez}}, \ and\ \bibinfo {author} {\bibfnamefont
  {A.}~\bibnamefont {Vuorinen}},\ }\href {\doibase 10.1103/PhysRevD.94.106008}
  {\bibfield  {journal} {\bibinfo  {journal} {Phys. Rev. D}\ }\textbf {\bibinfo
  {volume} {94}},\ \bibinfo {pages} {106008} (\bibinfo {year} {2016})},\
  \Eprint {http://arxiv.org/abs/1609.03480} {arXiv:1609.03480 [hep-th]}
  \BibitemShut {NoStop}%
\bibitem [{\citenamefont {Ecker}\ \emph {et~al.}(2017)\citenamefont {Ecker},
  \citenamefont {Hoyos}, \citenamefont {Jokela}, \citenamefont
  {Rodr\'{i}guez~Fern\'{a}ndez},\ and\ \citenamefont
  {Vuorinen}}]{Ecker:2017fyh}%
  \BibitemOpen
  \bibfield  {author} {\bibinfo {author} {\bibfnamefont {C.}~\bibnamefont
  {Ecker}}, \bibinfo {author} {\bibfnamefont {C.}~\bibnamefont {Hoyos}},
  \bibinfo {author} {\bibfnamefont {N.}~\bibnamefont {Jokela}}, \bibinfo
  {author} {\bibfnamefont {D.}~\bibnamefont {Rodr\'{i}guez~Fern\'{a}ndez}}, \
  and\ \bibinfo {author} {\bibfnamefont {A.}~\bibnamefont {Vuorinen}},\ }\href
  {\doibase 10.1007/JHEP11(2017)031} {\bibfield  {journal} {\bibinfo  {journal}
  {JHEP}\ }\textbf {\bibinfo {volume} {11}},\ \bibinfo {pages} {031} (\bibinfo
  {year} {2017})},\ \Eprint {http://arxiv.org/abs/1707.00521} {arXiv:1707.00521
  [hep-th]} \BibitemShut {NoStop}%
\bibitem [{\citenamefont {Ishii}\ \emph {et~al.}(2019)\citenamefont {Ishii},
  \citenamefont {Jarvinen},\ and\ \citenamefont {Nijs}}]{Ishii:2019gta}%
  \BibitemOpen
  \bibfield  {author} {\bibinfo {author} {\bibfnamefont {T.}~\bibnamefont
  {Ishii}}, \bibinfo {author} {\bibfnamefont {M.}~\bibnamefont {Jarvinen}}, \
  and\ \bibinfo {author} {\bibfnamefont {G.}~\bibnamefont {Nijs}},\ }\href
  {\doibase 10.1007/JHEP07(2019)003} {\bibfield  {journal} {\bibinfo  {journal}
  {JHEP}\ }\textbf {\bibinfo {volume} {07}},\ \bibinfo {pages} {003} (\bibinfo
  {year} {2019})},\ \Eprint {http://arxiv.org/abs/1903.06169} {arXiv:1903.06169
  [hep-ph]} \BibitemShut {NoStop}%
\bibitem [{\citenamefont {Heller}\ and\ \citenamefont
  {Spalinski}(2015)}]{Heller:2015dha}%
  \BibitemOpen
  \bibfield  {author} {\bibinfo {author} {\bibfnamefont {M.~P.}\ \bibnamefont
  {Heller}}\ and\ \bibinfo {author} {\bibfnamefont {M.}~\bibnamefont
  {Spalinski}},\ }\href {\doibase 10.1103/PhysRevLett.115.072501} {\bibfield
  {journal} {\bibinfo  {journal} {Phys. Rev. Lett.}\ }\textbf {\bibinfo
  {volume} {115}},\ \bibinfo {pages} {072501} (\bibinfo {year} {2015})},\
  \Eprint {http://arxiv.org/abs/1503.07514} {arXiv:1503.07514 [hep-th]}
  \BibitemShut {NoStop}%
\bibitem [{\citenamefont {Heller}\ \emph {et~al.}(2018)\citenamefont {Heller},
  \citenamefont {Kurkela}, \citenamefont {Spali\'nski},\ and\ \citenamefont
  {Svensson}}]{Heller:2016rtz}%
  \BibitemOpen
  \bibfield  {author} {\bibinfo {author} {\bibfnamefont {M.~P.}\ \bibnamefont
  {Heller}}, \bibinfo {author} {\bibfnamefont {A.}~\bibnamefont {Kurkela}},
  \bibinfo {author} {\bibfnamefont {M.}~\bibnamefont {Spali\'nski}}, \ and\
  \bibinfo {author} {\bibfnamefont {V.}~\bibnamefont {Svensson}},\ }\href
  {\doibase 10.1103/PhysRevD.97.091503} {\bibfield  {journal} {\bibinfo
  {journal} {Phys. Rev. D}\ }\textbf {\bibinfo {volume} {97}},\ \bibinfo
  {pages} {091503} (\bibinfo {year} {2018})},\ \Eprint
  {http://arxiv.org/abs/1609.04803} {arXiv:1609.04803 [nucl-th]} \BibitemShut
  {NoStop}%
\bibitem [{\citenamefont {Romatschke}(2018)}]{Romatschke:2017vte}%
  \BibitemOpen
  \bibfield  {author} {\bibinfo {author} {\bibfnamefont {P.}~\bibnamefont
  {Romatschke}},\ }\href {\doibase 10.1103/PhysRevLett.120.012301} {\bibfield
  {journal} {\bibinfo  {journal} {Phys. Rev. Lett.}\ }\textbf {\bibinfo
  {volume} {120}},\ \bibinfo {pages} {012301} (\bibinfo {year} {2018})},\
  \Eprint {http://arxiv.org/abs/1704.08699} {arXiv:1704.08699 [hep-th]}
  \BibitemShut {NoStop}%
\bibitem [{\citenamefont {Denicol}\ and\ \citenamefont
  {Noronha}(2020)}]{Denicol:2019lio}%
  \BibitemOpen
  \bibfield  {author} {\bibinfo {author} {\bibfnamefont {G.~S.}\ \bibnamefont
  {Denicol}}\ and\ \bibinfo {author} {\bibfnamefont {J.}~\bibnamefont
  {Noronha}},\ }\href {\doibase 10.1103/PhysRevLett.124.152301} {\bibfield
  {journal} {\bibinfo  {journal} {Phys. Rev. Lett.}\ }\textbf {\bibinfo
  {volume} {124}},\ \bibinfo {pages} {152301} (\bibinfo {year} {2020})},\
  \Eprint {http://arxiv.org/abs/1908.09957} {arXiv:1908.09957 [nucl-th]}
  \BibitemShut {NoStop}%
\end{thebibliography}%

\end{document}